\pdfoutput=1
\documentclass[useAMS,usenatbib]{mn2e}
\usepackage{amsmath,amssymb,graphicx,subfigure}
\usepackage{xcolor}

\newcommand{\msun}{\ensuremath{\,\mathrm{M}_\odot}}	% msun in text
\newcommand{\iso}[2] {$^{#1}{\rm #2}$}		% isotope in text
\newcommand{\miso}[2] {^{#1}{\rm #2}}		% isotope in math
\title[Evolution and nucleosynthesis in massive stars]
{Code dependencies of pre-supernova evolution and\\
nucleosynthesis in massive stars: Evolution to the end of\\
core helium burning}
\author[S. W. Jones et al.]{S. Jones$^{1,2,9}$\thanks{Email: s.w.jones@keele.ac.uk}, R. Hirschi$^{2,3,9}$,
 M. Pignatari$^{4,9}$, A. Heger$^{5,6,7,9}$, C. Georgy$^{2}$,
\newauthor
N. Nishimura$^{2,9}$, C. Fryer$^{8,9}$, F. Herwig$^{1,7,9}$
\\
$^1$ Department of Physics and Astronomy, University of Victoria, BC V8W 3P6, Canada \\
$^2$ Astrophysics Group, Lennard Jones Building, Keele University ST5
5BG, UK \\
$^3$ Kavli Institute for the Physics and Mathematics of the Universe (WPI), The University of Tokyo, Kashiwa, Chiba 277-8583, Japan \\
$^4$ Department of Physics, University of Basel, Klingelbergstrasse 82, CH-4056 Basel, Switzerland \\
$^5$ Monash Centre for Astrophysics,
School of Mathematical Sciences,
Building 28, M401,
Monash University, Vic 3800, Australia \\
$^6$ Minnesota Institute for Astrophysics,
School of Physics \& Astronomy, University of Minnesota,
Minneapolis, MN 55455, U.S.A. \\
$^7$ Joint Institute for Nuclear Astrophysics, University of Notre Dame, IN 46556, USA \\
$^8$ Computational Physics and Methods (CCS-2), LANL, Los Alamos, NM, 87545, USA \\
$^9$ NuGrid Collaboration, http://www.nugridstars.org}
\begin{document}

\date{Received: 10-NOV-2014; accepted: 12-DEC-2014.}

\pagerange{\pageref{firstpage}--\pageref{lastpage}} \pubyear{2012}

\maketitle

\label{firstpage}

% !TEX root = ./ms.tex

\begin{abstract}
Massive stars are key sources of radiative, kinetic, and chemical
feedback in the universe.  Grids of massive star models computed by
different groups each using their own codes, input physics
choices and numerical approximations, however, lead to
inconsistent results for the same stars.  We use three of these 1D
codes---GENEC, KEPLER and MESA---to compute non-rotating stellar
models of $15\msun$, $20\msun$, and $25\msun$ and compare their
nucleosynthesis.  We follow the evolution from the main sequence until
the end of core helium burning.  The GENEC and KEPLER models hold
physics assumptions used in large grids of published models.  The MESA
code was set up to use convective core overshooting such that the CO
core masses are consistent with those obtained by GENEC.  For all
models, full nucleosynthesis is computed using the NuGrid
post-processing tool MPPNP.

We find that the surface abundances predicted by the models are in
reasonable agreement.  In the helium core, the standard deviation of
the elemental overproduction factors for Fe to Mo is less than
$30\,\%$---smaller than the impact of the present nuclear physics
uncertainties.  For our three initial masses, the three stellar evolution
codes yield consistent results.  Differences in key properties
of the models, e.g., helium and CO core masses and the time spent as
a red supergiant, are traced back to the
treatment of convection and, to a lesser extent, mass loss.  The
mixing processes in stars remain the key uncertainty in stellar
modelling.  Better constrained prescriptions are thus necessary to
improve the predictive power of stellar evolution models.

\end{abstract}


% !TEX root = ./ms.tex

\section{Introduction}
\label{intro}

Over the last few decades, several groups have calculated and
published large grids of models across the massive-star mass range and
spanning several initial metallicities.  Such grids of models have
proved invaluable for those wishing to simulate, for example,
core-collapse supernovae
\citep[e.g.,][]{Oconnor2011,Mueller2012a,Ugliano2012,Couch2013,Nakamura2014},
galactic chemical evolution
\citep[e.g.,][]{Chiappini1997,Kawata2003,Cescutti2014}, or population
synthesis \citep[e.g.,][]{Bruzual2003,Eldridge2009}.  They are also
important resources with which observations of directly-imaged
supernova progenitors can be compared
\citep[e.g.,][]{Smartt2009,Maund2011,Fraser2011}.  It is difficult to
quantify the uncertainties in predictive simulations or in determining
the nature of an observation when the uncertainties in the underlying
stellar evolution and nucleosynthesis models are themselves rather
elusive.  For such studies, it would be advantageous if one was able
to know a priori some of the qualitative and quantitative differences
that a simulation using the results of stellar modelling would exhibit
had the stellar models been computing using different assumptions or
indeed with a different stellar evolution code.

Massive stars are those which produce an inert iron core and,
ultimately, explode as core-collapse supernovae \citep[see, for a
  review,][]{Woosley2002zz}.  At solar metallicity, massive stars are
those with initial masses greater than about $8\msun$--$10\msun$, just
below which super-AGB stars and the progenitors of electron-capture
supernovae are formed
\citep{Nomoto1984,EldridgeTout2004,Poelarends2008,Jones2013}.  Above
initial masses of roughly 25\msun\ massive stars forming inert iron
cores may end their lives as weak or failed supernovae in which black
holes are formed \citep{Heger2003}.  Dim, weakly energetic and failed
supernovae, however, are likely the result of more complicated details
of the stellar evolution than simply the initial mass of the star
\citep{Oconnor2011,Ugliano2012}.  Those more massive still (around
120\msun\ and higher) can become unstable due to electron-positron
pair creation before the star can develop an iron core
\citep[e.g.,][]{Heger2002,Heger2003,Yusof2013}.  These stars will
  likely not leave behind a compact remnant at all.  In this study, we
  focus on the main massive star range studied, between 15\msun and
  25\msun.

\citet{Martins2013} have recently compared the observable properties
predicted by massive star models computed with the MESA, STAREVOL,
GENEC, STERN, Padova, FRANEC codes to Galactic observations.
Their study shows that the post main sequence evolution differs
significantly between the different codes but it is hard to analyse
the differences when so many input physics are different between the
different stellar evolution codes.  In this study, by using
a post-processing method for the full nucleosynthesis, we eliminate
differences that could have arisen from different codes using
different rates.
Thus, this study focuses on a few key ``stellar'' (mainly the
treatment of convection) and ``numerical'' ingredients of the
models.  For this reason, we also do not consider the effects of
rotation in this study.

Massive stars are the primary producers of \iso{16}{O} in the
universe, along with most of the alpha elements (e.g., Mg, Si, Ca, and
Ti) and a significant fraction of the Fe-group elements
\citep[][]{kobayashi:11}.  Therefore, galactic chemical evolution (GCE)
simulations need to rely on
robust predictions for massive stars, over a large range of stellar
masses and initial metallicities.  Massive stars are also the host of
the weak \textsl{s}-process, responsible for the majority of the
\textsl{s}-process abundances with $60<A<90$ in the Solar System
\citep[e.g.,][]{Raiteri1993,the:07}.  This includes most
of the solar Cu, Ga, and Ge \citep{Pignatari2010}.
\citet{Tur2007,Tur2009,West2013} have investigated the sensitivity of
weak \textsl{s}-process production in massive stars to both the
triple-$\alpha$ and \iso{12}{C}($\alpha,\gamma$)\iso{16}{O} reaction
rates.  The impact of the \iso{12}{C}($\alpha,\gamma$)\iso{16}{O} was
also considered in a number of different works, showing sometimes
different results \citep[see e.g.,][]{imbriani:01,eleid:04}.  The
uncertainties in these reaction rates, which are important during
helium- and carbon-burning, were shown by Tur et al. to induce large
changes in the remnant (proto-neutron star) masses, which propagate to
the final (explosive) nucleosynthesis yields of massive stars.
Uncertainties in the \iso{12}{C}+\iso{12}{C} reaction rate also
propagate through into uncertainties in weak \textsl{s}-process
element production, primarily via the impact that enhancing or
reducing the rate has on the stellar structure during carbon burning
\citep{gasques:07,Bennett2012,Pignatari2013}.

The aim of this work is to examine the structural differences in the
evolution of massive stars as predicted by different published and
unpublished stellar evolution models.  In addition, we study the
impact of these structural differences on the nucleosynthesis
processes of $15\msun$, $20\msun$ and $25\msun$ stars.
The NuGrid tools enable us to compare the nucleosynthesis in these
stellar models using the same set of reaction rates in a post-processing
mode, drawing the focus of the comparison to differences in the
structural evolution of the models and their impact on the nucleosynthesis.
There are several well-established codes able to compute the evolution of
massive stars: (in no particular order) GENEC, KEPLER, STERN
\citep{Brott2011}, STARS \citep{EldridgeTout2004}, FRANEC \citep{Limongi2012},
TYCHO \citep{Young2005}, and MESA to name but a few.
In this work, we limited ourselves to using the GENEC, KEPLER and MESA codes
(see Section~\ref{methodsec} for detailed descriptions of these codes)
as a representative sample in order to be able to compare codes in greater detail.

This paper is structured as follows: In Section~\ref{methodsec} we
describe in detail the physics assumptions and numerical
implementations of the codes. In Section~\ref{evol_section} the
structural evolution of the stellar models from the pre-main sequence
are described, compared and contrasted.  The nucleosynthesis
calculations are presented in Section~\ref{nucleo} and in
Section~\ref{discussion} we summarise our results and discuss their
implications in a broader context.

% !TEX root = ./ms.tex

\section[]{Methodology and input physics}
\label{methodsec}

Massive star models with initial masses of 15\msun, 20\msun, and
25\msun\ and initial metallicity $Z = 0.02$ were computed using three
different stellar evolution codes.  The calculations were performed
using the Geneva stellar evolution code \citep[GENEC hereafter;
  see][]{Eggenberger2008a,Ekstrom2012a}, KEPLER \citep[][and citations
  therein]{WZW78,Rauscher2002,WH07} and MESA
\citep[][]{MESA2011,MESA2013}, revision 3709.
Concerning the models computed for this study,
note that we do not expect any major changes between using
this revision of MESA or later revisions including updates as described
in \citet{MESA2013}.
Baseline stellar physics assumptions are made, and feedback
from rotation and magnetic fields is not considered as explained
in the Introduction.  In this section, we give a brief description of the codes and
list the main input physics and assumptions.

The goal is not to obtain the same answer with all three codes.
Rather, we aim to discuss the qualitative and quantitative differences
between results from the three codes using standard choices for input
physics for the GENEC and KEPLER codes, which have been adopted in
most of their respective published models.  Thus, we have focused on
the most-studied massive star mass range (15--25\msun).  The input physics choices for MESA are
explained presently in this section.  We hope that this study will
encourage the community to be vigilant when using grids of stellar
evolution models (and yields) published by the different groups and to
be better able to relate the differences in models to either the
different choices made for input physics or the different design of
the stellar evolution codes.

\subsection{Code background}
The Geneva Stellar Evolution Code is a longstanding code that is most
actively used to compute the structure and evolution of massive and
solar-type stars.
In its latest developments, it includes prescriptions for both
rotation and magnetic fields \citep{Eggenberger2008a,Ekstrom2012a}.
In GENEC, the equations of stellar structure, nuclear burning and
mixing are solved in a decoupled manner.  The structure equations are
solved by means of a relaxation method that is usually referred to as
the Henyey method \citep{Henyey1964a}.  In this time-implicit method
firstly the structure equations, followed by the nuclear burning and
finally the mixing are calculated in turn in an iterative scheme until
the desired precision is reached.

The very external layers of the star---the $2\,\%$ of the total mass
below the surface in non-rotating models---are not computed in the
same way as the interior in GENEC.  These external layers are solved
using the pressure as the independent variable instead of the mass
coordinate, allowing for a better discretisation of the
equations.  Moreover, partial ionisation is accounted for in the
equation of state (EOS).  For these external layers, GENEC assumes
safely that there is no energy generation from nuclear burning.

The \emph{Modules for Experiments in Stellar Astrophysics} (MESA,
\citealp{MESA2011,MESA2013}) program \texttt{MESA/star} is designed to
solve the equations of stellar structure in a fully coupled manner.
It is important to note that in MESA, different studies may use
different choices for input physics (e.g., either the Ledoux or
Schwarzschild criterion for convective stability) and there is
not a single \emph{recommended} set of input physics, although a
time-dependent exponentially-decaying diffusion scheme is typically
used for convective boundary mixing.  For this study, as described
below, we chose input physics parameters for MESA that are as similar
as possible to those in GENEC (e.g., mass loss prescriptions) and we
chose a convective boundary mixing $f$ parameter that produces a CO
core mass close to the CO core mass obtained with the GENEC and KEPLER
codes.  Thus, the MESA models represent something that a typical user
could readily reproduce without the need to modify the code.

Finally, KEPLER is a spherical symmetric implicit
hydrodynamic code \citep{WZW78} tuned to problems in stellar
evolution, with particular emphasis on proper modelling of the
advanced stellar evolution stages until onset of core collapse; the
evolution is stopped well before neutrino trapping starts to become
important.
Nuclear burning is implicitly coupled to the structure and full energy
conservation is assured.  Mixing is decoupled and treated in operator
split, however, time-dependent mixing in diffusion approximation is
used throughout.  Extended nuclear burning is followed separately in
co-processing from the zero-age main sequences to pre-SN using an
adaptive nuclear reaction network that automatically adds and removes
isotopes as needed \citep{Rauscher2002}.
KEPLER also includes rotation \citep{HLW2000} and transport processes
due magnetic fields \citep{Spruit2002,HWS2005} but these are not used
in the present study.  KEPLER is able to model both hydrodynamic
evolution phases and the supernova explosion (in a parametric way;
\citealt{Rauscher2002}).

\begin{table*}
\begin{center}
\begin{tabular}{p{4cm} p{4cm} p{4cm} p{4cm}}
\hline \hline
Code & MESA & GENEC & KEPLER \\
\hline \hline
\vspace{3pt} \\
Operator coupling & Fully coupled \vfill (structure+burn+mix) &
    Decoupled \vfill (structure, burn, mix) &
    Partially coupled \vfill (structure+burn, mix) \\
\vspace{3pt} \\
Mixing strategy & Schwarzschild criterion with exponential-diffusive
    convective boundary mixing & Schwarzschild criterion with penetrative
    overshooting & Ledoux criterion with fast semiconvection \\
\vspace{3pt} \\
Implementation of mixing & Diffusion approximation & Instantaneous up to oxygen burning, then diffusion approximation & Diffusion approximation \\
\vspace{3pt} \\
\hline \hline
\end{tabular}
\caption{Overview of the mixing assumptions and operator coupling
in the three stellar evolution codes (MESA, GENEC and KEPLER) that
were used in this work. All three codes include prescriptions for
rotation and magnetic fields, however these physics were not included
in the present study.}
\label{tab:code_overview}
\end{center}
\end{table*}

\subsection{Input physics}
\label{sec:inputphysics}
There are numerous differences in the input physics assumptions made
by the three stellar evolution codes that were used in this work.
It is critical when comparing the results produced with different codes
to begin with a good understanding of their fundamental differences.
To that end, in this section we list, compare and contrast the input physics
assumptions made concerning the initial composition, opacities, nuclear
reaction networks, nuclear reaction rates, mass loss prescriptions, equation
or state, convection and overshooting and, finally, the initial models and
treatment of the pre-main sequence.

\subsubsection{Initial composition}
The initial elemental abundances are scaled to $Z=0.02$ from the solar
distribution given by \cite{GrevesseNoels1993}\footnote{we acknowledge
  the measurements by \citet{GrevesseNoels1993} have since been
  succeeded by \citet{Asplund2009}, however, this will not change
  dramatically any of the results of this code comparison study.}  and
the isotopic percentage for each element is given by
\citet{Lodders2003}.

\subsubsection{Opacities}
The initial composition corresponds directly to
the OPAL Type 2 opacity tables \citep{Iglesias_CO_1996} used in GENEC,
MESA and KEPLER (see below) for the present work.  The electron
conduction opacities used in GENEC are taken from \cite{Iben1975}.
For lower temperatures in MESA and GENEC, the corresponding opacities
from \citet{2005ApJ...623..585F} are used.  The MESA opacity tables
are in fact constructed from several sources, including the equations
of \citet{Buchler1976} for log$_{10}(T/\mathrm{K})>8.7$ where Compton
scattering becomes the dominant source of radiative opacity.  For
further details of the MESA opacity tables outside of the regions
discussed here, we refer the reader to \citet{MESA2011}.  The
opacities used in KEPLER are the same as in \citet{Rauscher2002} and
\citet{WH07}: for temperatures below $10^9\,$K a set of opacities also
described in \citet{AH1998PhDT} are used, with low-temperature opacity
tables from \citet{Alexander1994}, opacity tables from
\citet{Iglesias_CO_1996} at higher temperatures and for high
temperatures and enriched compositions the Los Alamos Opacity tables
\citep{Huebener1964} and electron conduction is included.  At higher
temperature the opacities as described in \citet{WZW78} are used,
including
\citet{Iben1975,Christy1966,Cooper1973,Chin1965,Sampson1959,HL1969,Canuto1970}
where applicable and where outside the tables above.

\subsubsection{Nuclear reaction network and reaction rates}
The nuclear reaction network is an essential supplement to the
structure equations, the equation of state and the opacities.
Because stellar evolution is so complex, it is costly to make a
detailed evaluation of the nuclear composition for the entire star
while solving the equations of stellar structure.  Despite this, full
yields from the stellar nucleosynthesis are highly desirable for
galactic chemical evolution to test our understanding of the
production sites of the heavy elements with $A\gtrsim60$.  In MESA, we
use a network of $171$ nuclear species (detailed in
Table~\ref{MESA_network_table}).
GENEC includes the main reactions for hydrogen and helium burning
phases and an alpha-chain type network for the advanced burning phases
\citep[even if it is now possible to extend it to a wide range of
  nuclear species and isotopes;][]{Frischknecht2012a}. The isotopes
included explicitly in the network are listed in
Table~\ref{MESA_network_table}.  Note that additional isotopes are
included implicitly to follow the pp chains, CNO tri-cycles and
$(\alpha,p)(p,\gamma)$ reactions in the advanced stages.  In KEPLER,
the APPROX-19 network follows a very similar approach, based on an
alpha-chain network with light isotopes added and including additional
reactions implictly, e.g., for the CNO cycle and conversion of
$^{22}$Ne to $^{24}$Mg.  This network is used for energy generation
and is implicitly coupled to structure, ensuring energy conservation.
KEPLER can also follow an extended adaptive nuclear reaction network
(BURN) in co-processing \citep[see][and references
  therein]{Rauscher2002,WH07} or fully coupled \citep{WH2004},
replacing the APPROX-19 network.  For the present work we have used
the first approach (APPROX-19 combined with BURN in co-processing) to
have results more consistent with the other codes.

In GENEC, most reaction rates were taken from the NACRE
\citep{AnguloNACRE1999} reaction rate compilation for the experimental
rates and from their
website\footnote{http://pntpm3.ulb.ac.be/Nacre/nacre.htm} for
theoretical rates, while in MESA, preference was given to the REACLIB
compilation \citep{Cyburt2010}.
This includes several rates from the NACRE compilation fitted with the
standard REACLIB fitting coefficients, for example for the
$\miso{22}{Ne}(\alpha,n)\miso{25}{Mg}$ reaction.
In KEPLER, reaction rates gererally
are from \citet{Rauscher00}, supplemented by rates from
\citeauthor{CaughlanFowler1988} (\citeyear{CaughlanFowler1988};
\citealt{Rauscher2002}).
There are a few exceptions concerning the key energy-producing
reactions.  In GENEC, the rate of \citet{Mukhamedzhanov2003} was used
for $^{14}$N(p ,$\gamma$)$^{15}$O below $0.1$ GK and the lower limit
NACRE rate was used for temperatures above $0.1$ GK.  This combined
rate is very similar to the more recent LUNA rate \citep{Imbriani2004}
at relevant temperatures, which is used in MESA.
In MESA and GENEC the \citet{Fynbo_3a_2005} rate was used for the
$3\alpha$ reaction and the \citet{Kunz2002} rate for
$^{12}$C($\alpha,\gamma$)$^{16}$O.  The \iso{12}{C}+\iso{12}{C} and
\iso{16}{O}+\iso{16}{O} reaction rates were those of
\citet{CaughlanFowler1988}.
In KEPLER we use the
$\miso{12}{C}(\alpha,\gamma)\miso{16}{O}$ rate of
\citet{Buchmann96,Buchmann97} multiplied by a factor 1.2
as suggested by \citet{West2013}, and
the rate of \citet{CaughlanFowler1988} for $3\alpha$.
In GENEC, the $^{22}$Ne($\alpha$, n)$^{25}$Mg rate was taken from
\citet{Jaeger2001} and used for $T \leq 1$ GK.  The NACRE rate was
used for higher temperatures.  The $\miso{22}{Ne}(\alpha, n)\miso{25}{Mg}$
rate competes with $^{22}$Ne($\alpha, \gamma$)$^{26}$Mg, where the
NACRE rate was used.
The $\miso{22}{Ne}(\alpha, n)\miso{25}{Mg}$ rate from NACRE
\cite{AnguloNACRE1999} was used in the MESA code.

\begin{table}
\begin{center}
\begin{tabular}{c c c c}
\hline \hline
Code & MESA & GENEC & KEPLER$^*$ \\
\hline
Element & A & A & A \\
\hline \hline
n&1&&1\\
p$^\dagger$&&&1\\
H & 1, 2 & 1 & 1\\
He & 3, 4 & 3, 4 & 3, 4\\
Li & 7 & &\\
Be & 7 & &\\
B & 8 & &\\
C & 12, 13 & 12, 13, & 12\\
N & 13 -- 15 & 14, 15 & 14 \\
O & 15 -- 21 & 16, 17, 18 & 16\\
F & 17 -- 23 & &\\
Ne & 18 -- 25 & 20, 21, 22 & 20\\
Na & 20 -- 27 & &\\
Mg & 22 -- 30 & 24, 25, 26 & 24\\
Al & 24 -- 31 & &\\
Si & 27 -- 32 & 28 & 28\\
P & 30 -- 34 & & \\
S & 31 -- 36 & 32 & 32\\
Cl & 33 -- 38 & & \\
Ar & 35 -- 40 & 36 & 36\\
K & 37 -- 42 & & \\
Ca & 39 -- 44 & 40 & 40\\
Sc & 41 -- 46 & & \\
Ti & 43 -- 50 & 44 & 44\\
V & 45 -- 52 & & \\
Cr & 47 -- 56 & 48 & 48\\
Mn & 49 -- 58 & & \\
Fe & 51 -- 60 & 52 & 52, 54\\
Co & 53 -- 62 & &\\
Ni & 55 -- 62 & 56 & 56\\
\hline \hline
\end{tabular}

\caption{Isotopes included in the nuclear reaction network of the
  various codes used in this paper
  \newline $^*$ APPROX-19 network \newline $^\dagger$
  protons from photo-disintegration treated separately from $^1$H in
  KEPLER for network stability.}

\label{MESA_network_table}
\end{center}
\end{table}

\subsubsection{Mass Loss}
\label{mass_loss_section}
Several mass loss rates are used depending on the effective
temperature, $T_{\rm eff}$, and the evolutionary stage of the star in
GENEC. For main sequence massive stars, where $\log T_{\rm eff} >
3.9$, mass loss rates are taken from \citet{Vink2001}.  Otherwise the
rates are taken from \citet{deJager1988}.  One exception is the
$15\msun$ model, for which the mass loss rates of \citet{deJager1988}
are used for the full evolution.  For lower temperatures ($\log T_{\rm
  eff} < 3.7$), however, a scaling law of the form
\begin{equation}
\dot{M} = - 1.479 \times 10^{-14} \times \left(\frac{L}{\mathrm{L}_{\sun}}\right)^{1.7}
\end{equation}
is used, where $\dot{M}$ is the mass loss rate in solar masses per
year, $L$ is the total luminosity and $L_{\sun}$ is the solar
luminosity \citep[see][and references therein]{Ekstrom2012a}.  In
MESA, we adopt several mass loss rates according to the scheme used in
\citet[][wind scheme {\tt Dutch} in the code]{Glebbeek2009}. For
effective temperatures of $\log T_{\rm eff} < 4$ it uses mass loss
rates according to \citet{deJager1988}.  For $\log T_{\rm eff} > 4$
there are two prescriptions that can be used, depending on the
abundance of hydrogen at the surface.  For $X_{\rm S}(^1{\rm
  H})>0.4$---a criterion satisfied by all of the models in this study
throughout their entire evolution---the rates of \citet{Vink2001} are
used.  For the mass range considered in this paper (15\msun--25\msun),
the mass loss prescription used in KEPLER is from
\citet{Nieuwenhuijzen1990}; see \citet{WH07} for more details.

\subsubsection{Equation of state}
The equation of state (EOS) used in GENEC is that of a mixture of an
ideal gas and radiation with pressure and temperature as the
independent variables, and accounts for partial degeneracy in the
interior during the advanced stages \citep[see][]{Schaller1992a}.  In
MESA, the EOS is in tabular form and is constructed from OPAL
\citep{RogersOPAL2002} tables and for lower temperatures, the SCVH
\citep{Saumon1995} tables. For intermediate conditions, these two
tables are blended in a pre-processing manner. Outside of the regions
covered by these tables in the density--temperature plane, the HELM
\citep{TimmesHELM2000} and PC \citep{PotekhinPC2010} equations of
state are employed, again being blended at the boundaries of the
tables.  The equation of state used in KEPLER is very similar to the
one used in MESA. It is based on the work of
\citet{Blinnikov1996,Blinnikov1998} and has been compared to HELM by
\citet{TimmesHELM2000}.

\begin{table}
\begin{center}
\begin{tabular}{r c c}
\hline \hline
Model & $\tau_\mathrm{H}/10^6\mathrm{yr}$ & $\tau_\mathrm{He}/10^5\mathrm{yr}$ \\
\hline \hline
G15 & 11.4 & 13.0  \\
K15 & 11.2 & 20.5 \\
M15 & 12.5 & 12.9 \\
average & $11.7\pm0.545$ (5\%) & $15.5\pm 3.58$ (23\%) \\
\hline
G20 & 7.97 & 8.67 \\
K20 & 8.24 & 12.0 \\
M20 & 8.68 & 8.44 \\
average & $8.30\pm0.294$ (4\%) & $9.71\pm 1.64$ (17\%) \\
\hline
G25 & 6.52 & 6.74 \\
K25 & 6.66 & 8.77 \\
M25 & 6.88 & 6.58 \\
average & $6.69\pm0.146$ (2\%) & $7.36\pm 0.996$ (14\%) \\
\hline \hline
\end{tabular}
\caption{Nuclear burning lifetimes of all the stellar models with
  average values and standard deviations.}
\label{lifetimes_table}
\end{center}
\end{table}

\subsubsection{Convection and Overshooting}
\label{conv_ovr_section}
In GENEC, convective stability is determined on the basis of the
Schwarzschild criterion. Convective mixing is treated as instantaneous
from hydrogen burning up to neon burning, where the composition across
a convection zone is mass-averaged.
The temperature gradient in the convective zones of the deep interior
is assumed to be the adiabatic one, i.e. $\nabla=\nabla_{\rm ad}$,
which is good to about one part in a million. The treatment of the
external convective zones is made according to the mixing length
theory with a mixing length parameter $\alpha_\text{MLT} = 1.6$, and
accounts for the non-adiabaticity of the convection for cool stars.
Overshooting is only included for hydrogen- and helium-burning cores,
where an overshooting parameter of $\alpha_{\rm OV} = 0.2 H_P$ is used
as in previous grids of non-rotating models \citep{Schaller1992a}.
The overshooting is implemented as an extension of the convective core
by $\alpha_{\rm OV}$ above the strict Schwarzschild boundary and the
overshooting region is considered to be part of the convective zone.
Thus, the overshoot region is always chemically homogenised with the
convective core.

Convection in MESA is treated at all times as a diffusive process,
employing the assumptions of mixing-length theory throughout the star
with mixing length parameter $\alpha_{\rm MLT}=1.6$, for which the
diffusion coefficient is reduced exponentially at the boundary between
convective and radiative layers as a function of radius
\citep{Freytag1996,Herwig1997},
\begin{equation}
 D=D_0\mathrm{exp}\bigg{(}-\frac{2z}{f_\mathrm{CBM}H_{P,0}}\bigg{)}.
 \label{exponentialovershoot}
\end{equation}
$D$ is the diffusion coefficient as a function of distance $z$ from
the boundary location and $f_\mathrm{CBM}$ is a free parameter, for
which we assume the value of $0.022$ above and below all convective
zones except for below convective shells in which nuclear burning is
taking place, where we use $f_\mathrm{CBM}=0.005$.  $D_0$ is the
diffusion coefficient, taken equal to the mixing length diffusion
coefficient value ($D_\mathrm{MLT}$) at a distance
$f_\mathrm{CBM}H_\mathrm{P,S}$ inside the convection zone from the
Schwarzschild boundary.  At this location inside the convective zone,
the pressure scale height is $H_\mathrm{P,0}$, while $H_\mathrm{P,S}$
is the pressure scale height at the Schwarzschild boundary.  This is
because the value of $D_0$ drops sharply towards zero at the
Schwarzschild boundary.  This treatment of convective boundary mixing
is held from the main sequence until the end of core He-burning.

KEPLER uses the Ledoux criterion for convection with an efficient
semiconvection and a small amount of overshooting at the boundaries of
convective regions to ensure numerical stability.  KEPLER does not,
however, use the Ledoux criterion formulated like so:
\begin{equation}
\nabla_{\rm rad}>\nabla_{\rm ad} +
(\varphi/\delta)\nabla_\mu\;,
\label{Ledoux}
\end{equation}
but instead uses a form for a generalised equation of state as
described in Appendix A of \citet{HWS2005}, and also includes
thermohaline convection \citep[e.g.,][]{WH2004}.  A detailed
description of the mixing physics, including the semiconvective
diffusion coefficient and the treatment of overshooting, can be found
in \citet{Sukhbold2013}.

\subsubsection{Initial models and treatment of the pre-main sequence}
\label{sec:initial_models}
In KEPLER, the initial model is set up as a polytrope with index $n=3$
and central density of $\rho=0.1~\mathrm{g\,cm}^{-3}$.  The APPROX-19 network
accounts for $^3$He and follows its burning in the pre-MS phase just as
is does during the rest of the evolution; other light isotopes such as
deuterium of Lithium are not considered explicitly.
Pre-main sequence in the GENEC is not considered in this work.
The structure is converged from an approximative ZAMS structure,
homogeneous in chemical composition. It requires a few tens of timesteps
to converge towards a stabilised ZAMS structure that is considered to
be reached once the centre of the star is depleted by 0.003 of its initial
H content (in mass fraction).
The initial models in MESA are $n=1.5$ polytropes with central temperatures
of $9\times10^6~\mathrm{K}$. The equation of state and the mixing length
theory routines are called iteratively by the Newton-Raphson solver to converge
the total mass of the star, with the central density as the independent variable
\citep[see][]{MESA2011}. The evolution loop including the complete (user-specified)
reaction network is then begun starting from the initial model.

\subsection{Nucleosynthesis post-processing tool (MPPNP)}
\label{sec:MPPNP}
We use the NuGrid\footnote{www.nugridstars.org} multi-zone
post-processing nucleosynthesis tool MPPNP \citep{NuGrid2013}
to calculate the evolution of the composition in all of the stellar
models in our comparison.  From every time step calculated by the
stellar evolution codes, MPPNP reads the thermodynamic trajectories
($T$ and $\rho$) for the entire star and with a reaction network of
$1088$ nuclear species performs a fully implicit Newton-Raphson
calculation to evolve the composition.  This nuclear burning step is
then followed by a mixing step that solves the diffusion equation
using diffusion coefficients from the stellar evolution
calculations.  For more details of the NuGrid post-processing tool
MPPNP and the reaction rate compilations that we use, we refer the
reader to \citet{Bennett2012} and \citet{Pignatari2013}.  We used the
same key energy-producing reaction rates as are used in the MESA and
GENEC stellar evolution calculations to ensure consistency where
possible.

% !TEX root = ./ms.tex

\section{Structural Evolution}
\label{evol_section}
In this section, we describe differences in the stellar models arising
from the different physics assumptions and numerical implementation of
the codes, which will later be connected to differences in the
nucleosynthesis.
We separate the description of the hydrogen- and helium-burning
evolution in this section into two parts, concerning the interior
(\ref{early_cores}) and the surface (\ref{surface_section}).

\subsection{Evolution of convective hydrogen- and helium-burning cores}
\label{early_cores}
\subsubsection{Core hydrogen burning}
During the main sequence evolution of a massive star, fusion of
hydrogen into helium in the convective core results in a reduced
opacity and increased mean molecular weight, $\mu$.  The increase in
$\mu$ leads to an increase in luminosity ($L\propto\mu^4$).
The outer layers expand, exerting less weight on the
core, which also experiences a decrease in pressure.  The reduction in
opacity and pressure dominate over the increase in luminosity during
the main sequence and because $\nabla_{\rm{rad}}\propto\kappa L_rP$,
the radiative temperature gradient decreases.  The adiabatic
temperature gradient on the other hand, remains more or less
unchanged.  As a result, the material at the edge of the core becomes
convectively stable and therefore the mass of the convective core
decreases during the main sequence lifetime of the star.  As the
convective core recedes in mass, it leaves above it a (convectively
stable) region of radially decreasing mean molecular weight, i.e.,
with $\nabla_{\!\!\mu}\equiv \partial\ln\mu/\partial\ln P>0$.

\begin{figure}
  \centering \includegraphics[width=1.\linewidth,clip=True,trim=0mm
    9mm 0mm 20mm]{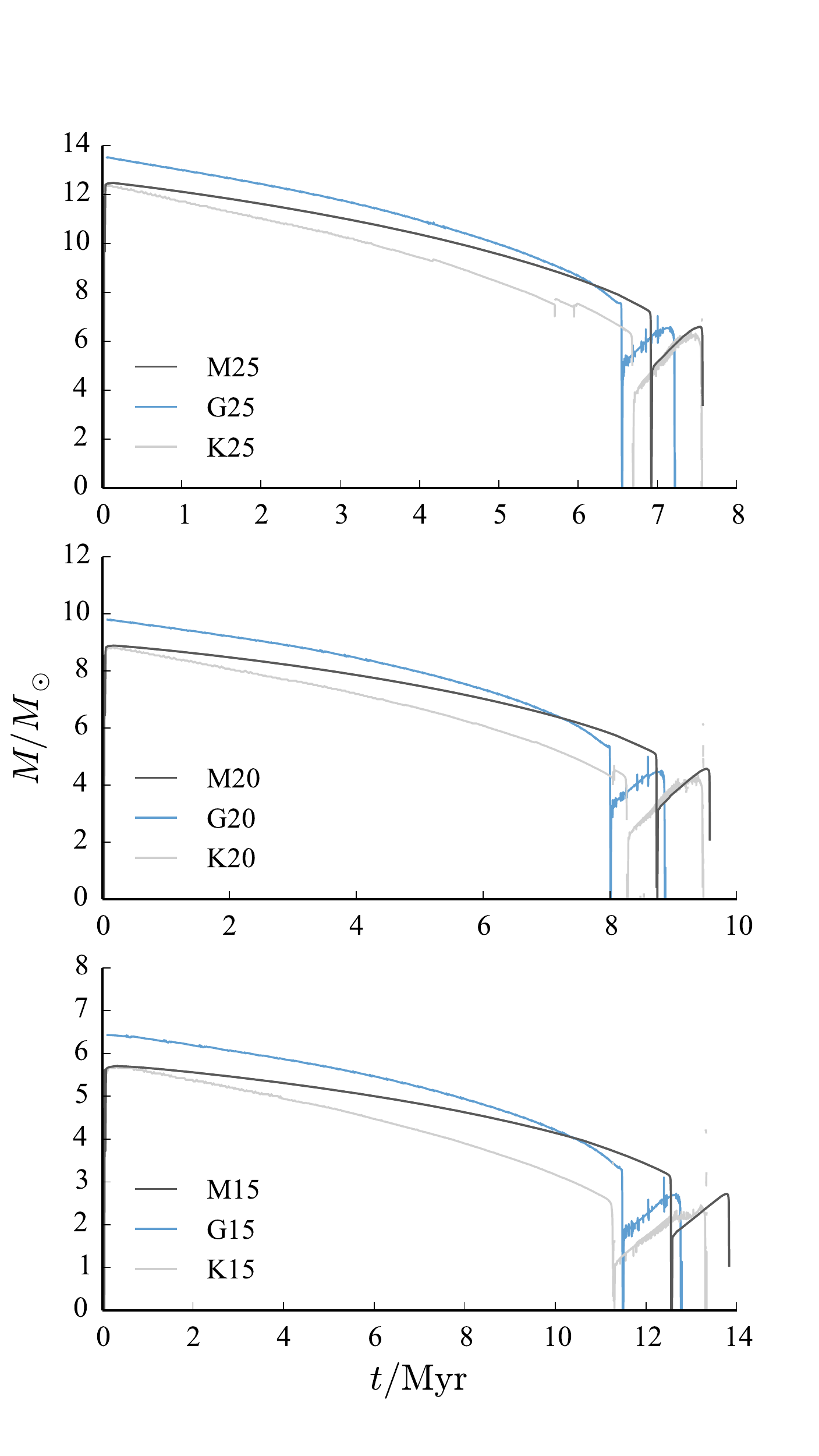}
  \caption{Convective core masses as a function of stellar age (Myr)
    during the core hydrogen- and helium-burning phases for
    the $15\msun$ (top panel), $20\msun$ (middle panel) and
    $25\msun$ (lower panel) models.}
  \label{hcore_t_plots}
\end{figure}

The time-evolution of the mass of the convective core is shown
in Fig.~\ref{hcore_t_plots} for the three masses we consider, computed
with the three stellar evolution codes (MESA, black curve; GENEC, blue curve;
KEPLER, silver curve).
The location of the boundary of the convective core at the ZAMS is not
dependent on the choice of convective stability criterion
(Schwarzschild or Ledoux), since the chemical composition of the star
is initially homogeneous.  This is confirmed by comparing the KEPLER
(Ledoux) and MESA (Schwarzschild) convective core masses at the
zero-age main sequence (ZAMS; Fig.~\ref{hcore_t_plots}).  There are,
however, differences between the MESA/KEPLER models and the GENEC
models already at the ZAMS due to the assumption of core
overshooting. Since GENEC assumes instantaneous, penetrative
overshooting, the overshoot region is always an extension of the
convective core that is effective immediately and thus its convective
cores are already more massive than in the other two codes at the
ZAMS.

As described earlier, in KEPLER convective stability is defined by the
Ledoux criterion and semiconvection is considered.  Although the
semiconvection is comparably fast \citep{WZW78,Sukhbold2013}, it is
still much slower than the mixing caused by either convection or
overshooting.  Thus in the GENEC models, which use the Schwarzschild
criterion with a penetrative overshooting of $0.2\,H_{\rm P}$, the
resulting convective cores are larger than in those calculated with
KEPLER for the entire duration of the main sequence.  The gradient
$dM_{\rm core}/dt$ is steepest in KEPLER because the gradient of
chemical composition $\nabla_{\!\!\mu}$ provides extra stability
against convection in the region above the core that was previously
part of the convective core.
In the MESA models, the Schwarzschild criterion is used to define
convective stability, as in GENEC, but overshooting is treated
diffusively with a diffusion coefficient that decays exponentially
into the radiative zone.  Not only does this result in different
extents of mixing in the two codes, but also in different time scales
of the extra mixing. In GENEC the convective region is assumed to be
instantaneously mixed during core H and He burning while MESA's
diffusive approximation accounts for the time-dependency of mixing.
$|dM_{\rm core}/dt|$ during the main sequence is thus smallest
(shallowest curve) for MESA because the time-dependent diffusive
treatment of overshooting mixes fresh fuel into the core more slowly
than the instantaneous mixing in GENEC.  Although the convective core
mass at the ZAMS is the same for the KEPLER and MESA models, the
diffusive treatment of overshoot mixing in MESA always results in a
larger convective core mass at the terminal-age main sequence (TAMS)
for the assumed value of $f_{\rm CBM}=0.022$ (see Eq.~\ref{exponentialovershoot}).

\begin{figure}
  \includegraphics[width=\columnwidth,clip=True,trim=0mm 9mm 0mm
    20mm]{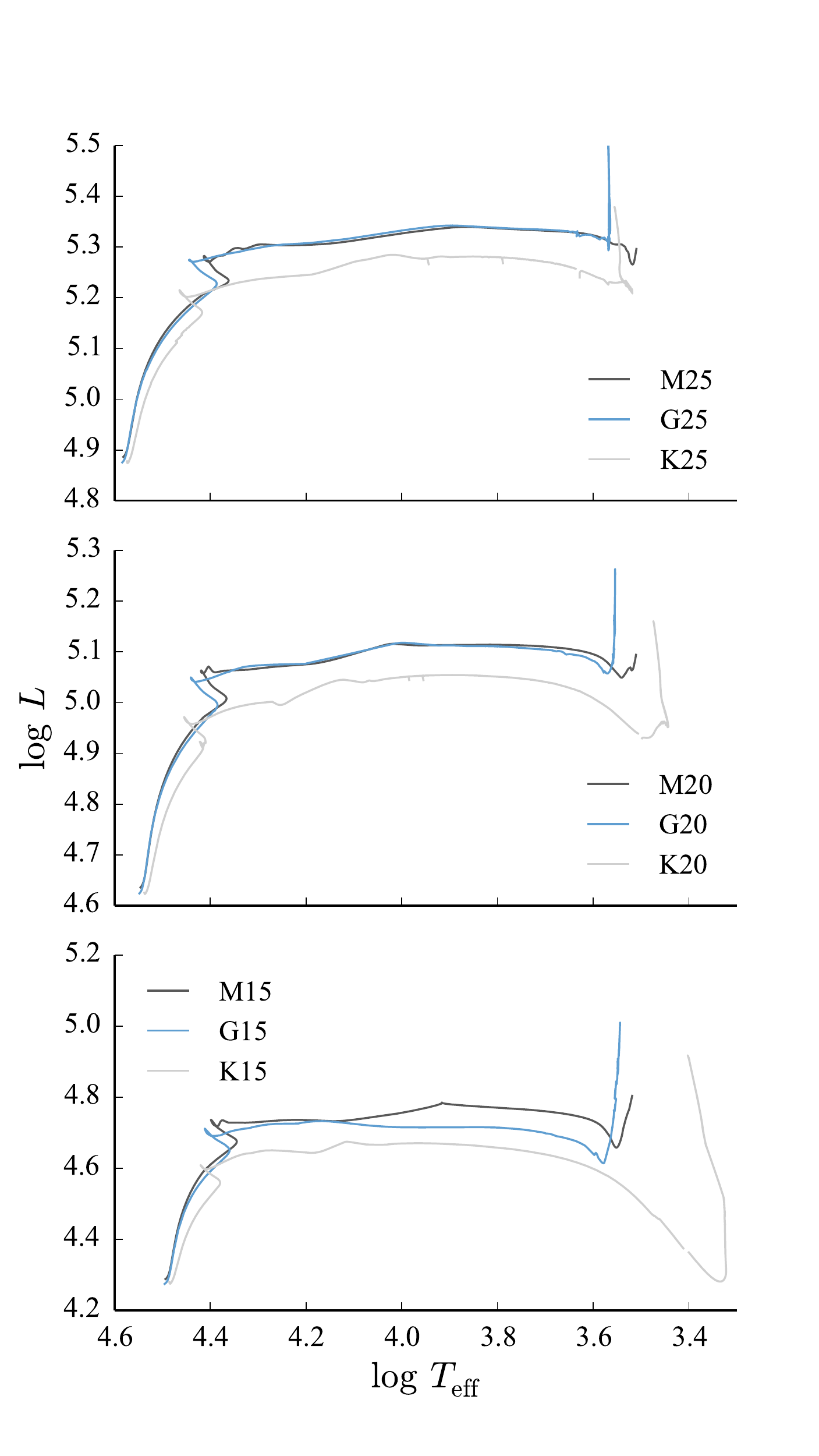}
  \caption{Evolution of the models in the Hertzsprung-Russell (H-R)
    diagram.}
  \label{HRDs}
\end{figure}
As a star evolves along the main sequence, its mean molecular weight
$\mu$ increases, as does its luminosity, $L$.  For a given initial
mass, the luminosities calculated by the KEPLER and GENEC codes at the
ZAMS agree very well (Fig.~\ref{HRDs}).  The ZAMS luminosities of the
MESA models are slightly higher because the convective core is
initially slightly more compact (smaller radius). This small
difference is likely due to the different treatment of the pre-MS
phase, which is described in Section~\ref{sec:initial_models} for the three
codes.

As the models evolve along the main sequence, the GENEC models
and MESA models have very similar luminosities for a given initial
stellar mass.  KEPLER, on the other hand, exhibits the lowest
luminosities during the main sequence after the ZAMS because of its
smaller cores.  The main sequence lifetime is determined by the
luminosity and the amount of fuel available. The hydrogen-burning
lifetimes in the GENEC and KEPLER models agree reasonably well, with
MESA always lasting longer on the main sequence.  This can be
attributed to the fact that although GENEC has higher luminosities
than KEPLER because of its larger convective cores, and so would burn
its fuel quicker, it also has more fuel available.  MESA, on the other
hand, exhibits slower growth in luminosity during the main sequence
than GENEC due to its diffusive treatment of overshooting.  Fuel is
constantly being mixed into the hydrogen-burning core in the MESA
models and as a result they show the longest main sequence lifetimes
of the three codes (see Table~\ref{lifetimes_table}; again, these
values are for $f_{\rm CBM}=0.022$).

\subsubsection{Core helium burning}
As described above and shown in Fig.~\ref{hcore_t_plots}, the KEPLER
models have consistently less massive convective cores at the TAMS
than the models calculated with the other two codes, which means that
they are less luminous throughout the core helium-burning phase and
thus have consistently longer helium-burning lifetimes (see
Table~\ref{lifetimes_table}).  The convective core of a massive star
grows in mass during the helium-burning lifetime (see
Fig.~\ref{hcore_t_plots}).  This is because the mass of the helium
(hydrogen-free) core is also growing in mass due to shell hydrogen
burning.  The core luminosity therefore increases and more helium-rich
material becomes convectively unstable.  Two other factors are the
increase of opacity due to the burning of helium to carbon and oxygen
and the density-sensitivity of the triple-alpha ($3\alpha$) reaction
rate.  The $3\alpha$ reaction rate has a second-order dependence on
the density while the $\miso{12}{C}(\alpha,\gamma)\miso{16}{O}$
reaction has only a first-order dependence. It is the latter reaction
that dominates the later part of helium burning.  It is the ingestion
of fresh helium into the late helium burning that significantly
reduces the central carbon-to-oxygen ratio at the end of central
helium burning and thereby has major impact on carbon burning and
beyond.  The C/O ratio in the core at the end of the helium-burning
phase is discussed in more detail in Section~\ref{sec:CO_ratio}.

In their longer helium-burning lifetimes, the convective core masses
in the KEPLER models have longer to grow and eventually become very
similar to those in GENEC by the point of central helium depletion
(Fig.~\ref{hcore_t_plots}).  In KEPLER the semiconvection is fast
enough to allow growth of the convective helium core whereas in plain
Ledoux convection often a split of the convective helium burning may
be found.  Note also that the semiconvection during hydrogen burning
in KEPLER can leave behind an extended region enriched in helium,
however, the first dredge-up may remove the outer part of this
enriched region.  An extended region enhanced in helium may allow the
hydrogen shell to grow faster than in other cases.
The MESA models develop the largest convective helium cores for all of
the three initial masses that we have considered, even though the MESA
models have the shortest helium-burning lifetimes (very closely
followed by GENEC).  This can be attributed to the more massive
hydrogen convective core at the TAMS in the MESA models.
The mass of the hydrogen and helium convective cores have important
implications for both the later evolutionary phases and the structure
at the pre-supernova stage, since they are strongly coupled to the
mass of the helium and CO cores ($M_\alpha$ and $M_{\rm CO}$,
respectively) at the pre-supernova stage.  Thus, the relationship
between the different codes for a given core mass ($M_\alpha$, $M_{\rm
  CO}$, etc.) at the pre-supernova stage is the same as is discussed
above (see Table~\ref{cores_table}), with the codes generally agreeing
to within a few percent.

\subsubsection{C/O ratio at the end of core helium burning}
\label{sec:CO_ratio}
In addition to the CO core mass, the carbon-burning evolution is very
sensitive to the C/O ratio at the end of the helium-burning phase.
The ratio of carbon to oxygen in the centre of all the models at the
time when the helium abundance first falls below $X_{\rm C}(^4{\rm
  He})=10^{-5}$ is shown in Table~\ref{CO_table}.  A lower C/O ratio
is indicative of higher temperatures and lower densities during core
helium burning---these conditions favour
\iso{12}{C}($\alpha$,$\gamma$)\iso{16}{O} over the 3$\alpha$ reaction
\citep{Woosley2002zz}.  As a result, the C/O ratio at the end of core
helium burning is lower for larger initial stellar mass.  Another
important factor is the treatment of convective boundary mixing. The
diffusive treatment of the extra mixing at the edge of the helium core
in the KEPLER and MESA models favours a lower C/O ratio. This is
because additional alpha particles can be mixed into the convective
core for the duration of the burning stage. In particular, the end of
core helium burning is especially sensitive.  A large proportion of
the alpha particles newly introduced into the core at this time will
react with carbon to produce oxygen, rather than reacting with other
alphas to produce carbon.  This can be attributed to the sensitivity
of the latter reaction to the cube of the alpha abundance. The
behaviour of the convective helium core as it approaches helium
exhaustion has been the subject of much discussion in the past.  In
particular, the presence of a `core breathing' phenomenon \citep[see,
  e.g.,][]{Castellani1985}. The buildup of \iso{12}{C} and more
critically \iso{16}{O} increases the opacity in the core and hence the
radiative temperature gradient also increases. In such a region at the
edge of the core there is also a steep gradient of the mean molecular
weight $\mu$.  The convectively stabilising effect of the $\mu$
gradient is overlooked by codes that adopt the Schwarzschild criterion
for convection, but not by those that adopt the Ledoux criterion and
consider the effects of semiconvective mixing.  The convective core
must thus grow to some extent in mass and thereby engulf fresh
\iso{4}{He}. The central abundances of \iso{4}{He}, \iso{12}{C} and
\iso{16}{O} are dramatically affected during such a `breathing pulse.'

\begin{figure}
  \centering
  \includegraphics[width=1.\linewidth,clip=True,trim=0mm
    10mm 0mm 20mm]{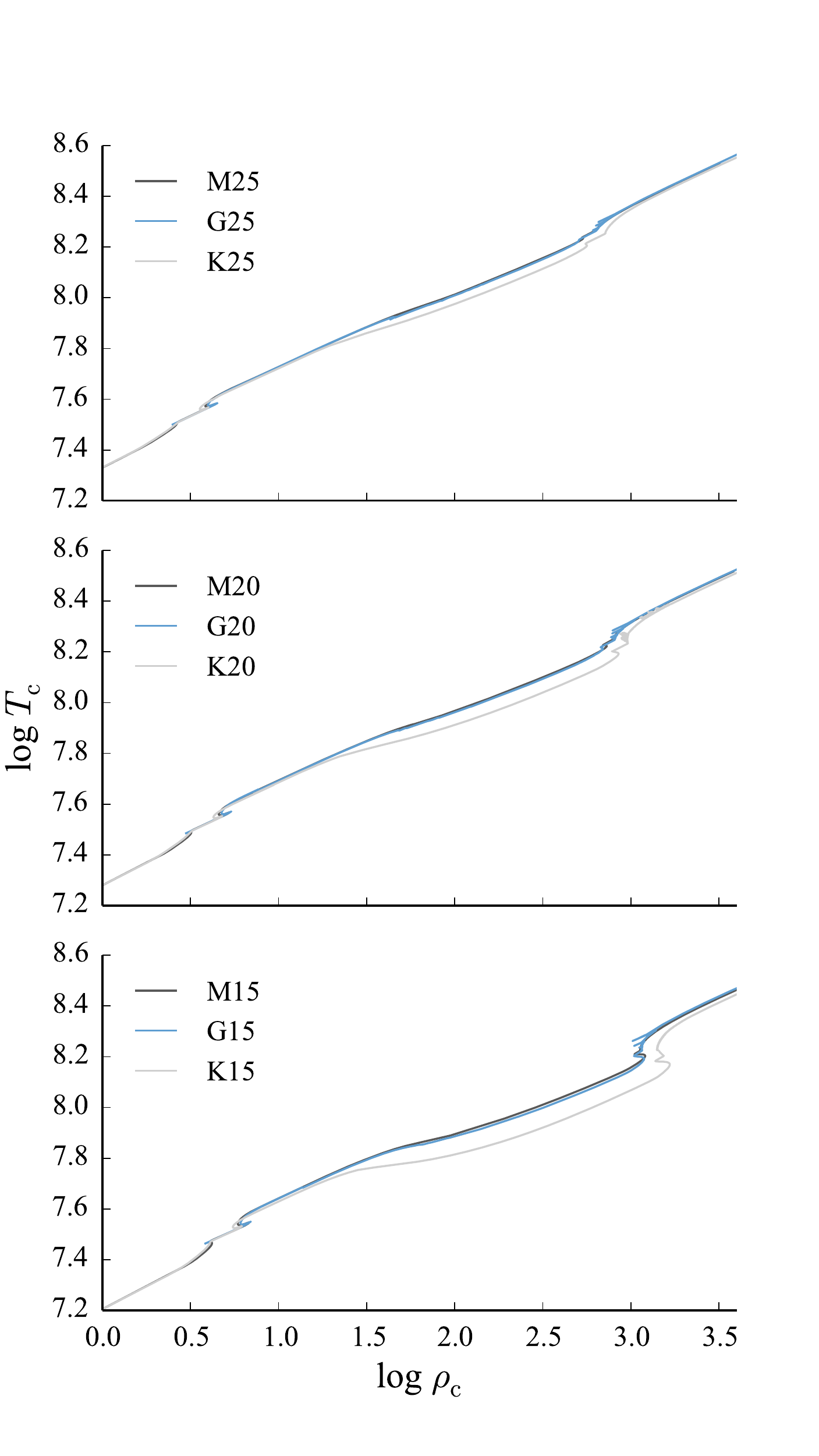}
  \caption{Evolution of the central temperature and density during the
  core hydrogen- and helium-burning phases.}
  \label{fig:tcrhoc}
\end{figure}

The decreasing trend of C/O ratio with initial mass is exhibited by
the models from all three codes.  This is true for both the inline
nuclear reaction networks and our full-network calculation in
post-processing mode (Table~\ref{CO_table}).  MESA models achieve
lower C/O than GENEC models for a given initial mass because of their
slightly more massive cores and diffusive overshooting (see discussion
above).  Another interesting factor that we had not initially expected
to affect the C/O ratio is the $\miso{22}{Ne}(\alpha,n)\miso{25}{Mg}$
reaction rate.  The \iso{22}{Ne} competes via this reaction with
\iso{12}{C} to capture the alpha particles.  It has in fact a large
enough effect that the C/O ratio will be sensitive to the
$\miso{22}{Ne}(\alpha,n)\miso{25}{Mg}$ reaction rate.  Of particular
note is a result that in the M20 model excluding the
$\miso{22}{Ne}(\alpha,n)$ and $\miso{22}{Ne}(\alpha,\gamma)$ reactions
results in a C/O ratio of $0.382$ as opposed to $0.395$ when the
reactions are included.  This ratio is important for energy generation
and the stellar structure during the post helium core burning
evolution.  Thus, \iso{22}{Ne} should be included in even small
nuclear reaction networks that are designed to calculate only the
energy generation inside the star.  The KEPLER models have the lowest
C/O ratio for a given mass out of the three codes. This is orthogonal
to the statement made earlier with regards to core mass, temperature,
density and the C/O ratio.  It is clear from the central temperature--density
evolution (Fig.~\ref{fig:tcrhoc})
that helium burns at lower temperature and higher density in the
KEPLER models than in the models from the other two codes.  Other
influencing factors are thus at play in the KEPLER models.  Firstly,
the $3\alpha$ and $\miso{12}{C}(\alpha,\gamma)\miso{16}{O}$ reaction
rates are different in the KEPLER models (see
Section~\ref{sec:inputphysics}).  Secondly, neither \iso{22}{Ne} or
\iso{25}{Mg} are included explicitly in the inline nuclear reaction
network of KEPLER.  We argue that this could compromise the accuracy
of the network by not explicitly following the competition between
\iso{22}{Ne} and \iso{12}{C} to capture the alpha particles that we
described above.  These two factors will influence the C/O ratio of
the inline network of KEPLER, however they should have no bearing on
the full network calculation that is done in post-processing mode.
Nevertheless, we do indeed find that the KEPLER models exhibit
consistently the lowest C/O ratios even in our post-processing
calculations.  This is because the core helium-burning lifetimes in
the KEPLER models are the longest. For example,
Table~\ref{lifetimes_table} shows that K15 burns helium in the core
for about 60\% longer than M15.  Artificially reducing the
helium-burning lifetime in the K15 model during the post-processing
stage indeed affects the C/O ratio.  When the helium-burning lifetime
of K15 is set to be the same as in M15, ${\rm C/O} = 0.468$, as
opposed to $0.347$ for the (longer) original lifetime.  This is still
a lower value than in G15, despite G15 having the higher temperature
and lower density that favour a lower C/O.  The sensitivity of the C/O
ratio to the implementation of overshoot mixing
(penetrative/diffusive) is highlighted by this simple test.

\begin{table}
\begin{center}
\begin{tabular}{r c c c}
\hline \hline
Model & M$_\mathrm{tot}/M_\odot$ & M$_\alpha/M_\odot$ & M$_\mathrm{CO}/M_\odot$ \\
\hline \hline
G15 & 12.13   &   4.79   &   2.86   \\
K15 & 10.77   &   3.94   &   2.64   \\
M15 & 12.15   &   4.76   &  2.99   \\
average & $11.69\pm0.65$ & $4.40\pm0.39$ & $2.83\pm 0.15$  \\
 & (6\%) &  (9\%) &  (5\%) \\
\hline
G20 & 13.97   &   6.83   &   4.54   \\
K20 & 13.11   &   5.99   &   4.38 \\
M20 & 15.40  &  6.77  & 4.65 \\
average & $14.16\pm 0.944$ & $6.53\pm 0.383$ & $4.52\pm 0.112$ \\
 & (7\%) &  (6\%) &  (2\%) \\
\hline
G25 & 13.74   &   9.19   &   6.48   \\
K25 & 12.34   &   8.14   &   6.28   \\
M25 & 12.82   &   9.13   &  6.82  \\
average & $12.97\pm 0.580$ & $8.82\pm 0.484$ & $6.53\pm 0.220$ \\
 & (4\%) &  (5\%) & (3\%) \\
\hline \hline
\end{tabular}
\caption{Total stellar mass (M$_\mathrm{tot}$) and masses of the helium
  (M$_\alpha$) and carbon-oxygen (M$_\mathrm{CO}$) cores
  at the end of core He-burning.}
\label{cores_table}
\end{center}
\end{table}

\subsection{Surface properties and HRD}
\label{surface_section}
\subsubsection{Evolution in the HRD}
At a given metallicity and initial mass, opacities and nuclear
reaction rates being the same, the evolution in the
Hertzsprung-Russell diagram (HRD) during the main sequence (MS) is
mostly determined by the evolution of the central convective
core\footnote{\footnotesize{In the mass range considered in this
    paper, the mass loss during the MS is weak enough not to influence
    significantly the tracks in the HRD.  This is not the case for
    higher mass models with $M>25\msun$.}}.  We have already discussed
the relationship between convective core mass and luminosity in the
section above for the main sequence.  For the reasons explained there
concerning the convective core, the KEPLER models exhibit a narrower
width of the main sequence and a lower turn-off luminosity
(Fig.~\ref{HRDs}).

\begin{figure}
  \centering \includegraphics[width=\columnwidth,clip=True,
  trim=0mm 9mm 0mm 20mm]{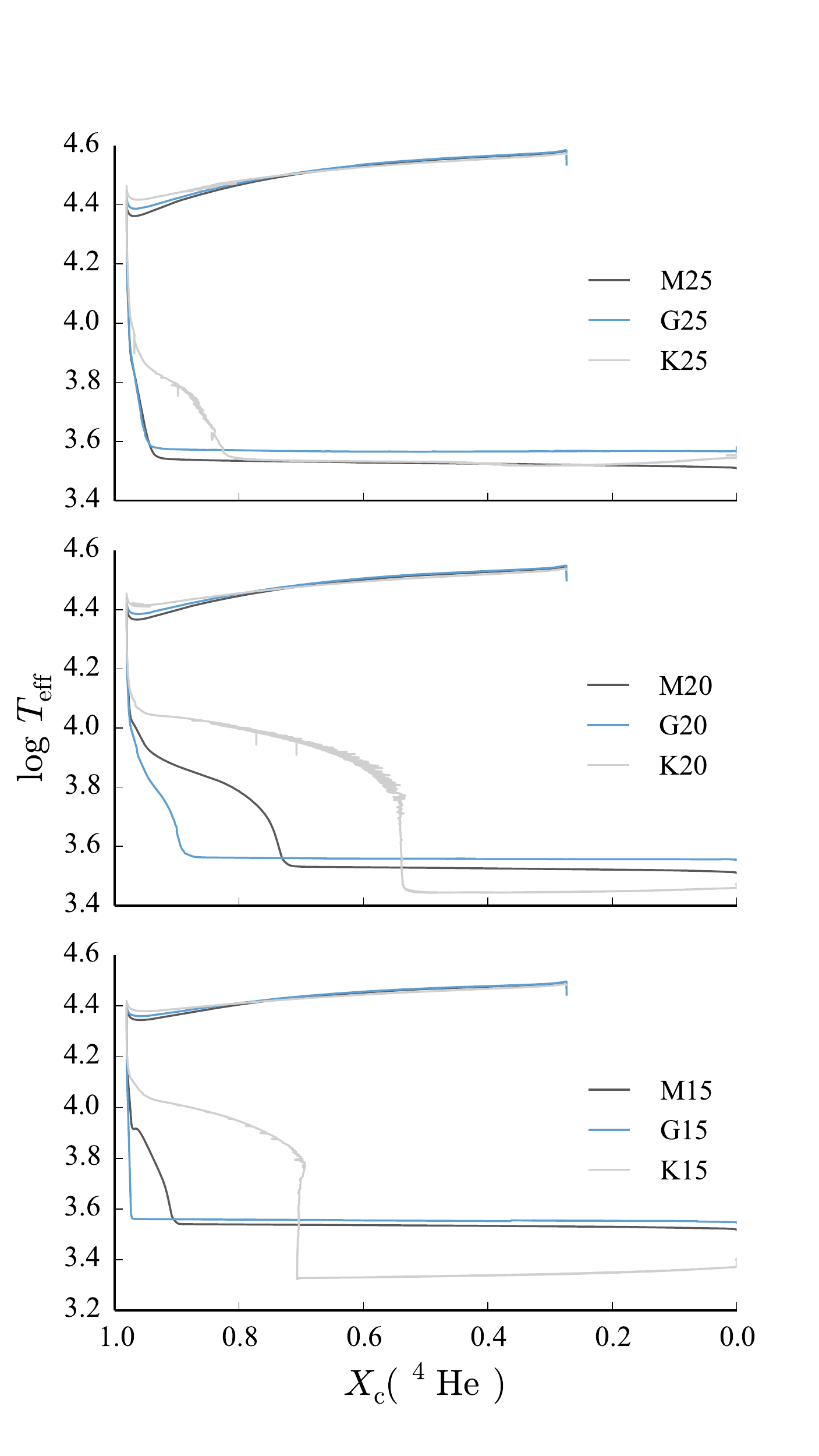}
  \caption{Effective temperature as a function of central
    $^4\mathrm{He}$ abundance for the $15\msun$ (bottom panel),
    $20\msun$ (middle panel) and $25\msun$ (top panel) models. In
    these diagrams, the evolution proceeds from the top-right corner
    to the top-left corner during the main sequence and from the left
    to bottom right corner during He-burning.}
  \label{teff_he4_plots}
\end{figure}

The trajectory of a stellar model in the H-R diagram after the main
sequence turn-off is determined by a complex interplay between the
helium core, the hydrogen-burning shell and the opacity of the
envelope.  In general, as the core contracts the envelope will
expand---a mirroring effect.  During the core helium-burning evolution
the structure is more complicated than during the main sequence. This
is because there is a hydrogen-burning shell between the core and the
envelope. The hydrogen-burning shell in fact provides a large fraction
of the stellar luminosity during core helium-burning.
Fig.~\ref{teff_he4_plots} shows effective temperature as a function of
central helium abundance for the models.

The KEPLER models of all three initial masses ignite helium burning
distinctly in the Hertzsprung gap. This only occurs in the $20\msun$
GENEC model while for MESA both the 20\msun\ and 25\msun\ models
undergo a blue helium ignition. GENEC still displays the reddest
helium ignition of the $20\msun$ models.  In the $20\msun$ MESA model,
shell hydrogen burning is initially very strong and a thick convection
zone in the hydrogen shell develops.  This provides a substantial
fraction of the star's luminosity and therefore the core requires a
slower rate of contraction in order to maintain hydrostatic
equilibrium than if such a strong hydrogen shell were not present. The
envelope expands less by the `mirror effect' and as a result the star
ignites helium before becoming a red supergiant. Out of all the KEPLER
models, the $20\msun$ model also develops the strongest convective
hydrogen shell.  At the TAMS, the KEPLER models have lower helium core
masses than the other codes -- the core contains a lower fraction of
the total mass of the star, and thus the mirror effect is less,
favouring a bluer helium-burning ignition than in the other codes.
This is a consequence of the use of the Ledoux criterion.  On the
other hand, the region above the core is more enriched in helium
allowing this more vigorous hydrogen shell burning.
\citet{Georgy2013grid} have shown that for rotating massive star
models, a hydrogen profile at the end of the main sequence that drops
steeply at the edge of the helium core favours the blue ignition of
helium.  In accord, a shallow hydrogen profile above the helium core
at the end of the main sequence favours a quicker evolution to the RSG
phase.  This is because with an extended hydrogen profile the hydrogen
shell will migrate outward in mass more rapidly and thus the core mass
will increase faster, leading to its contraction and hence, to the
expansion of the envelope.  These differences arose in models that had very similar
structures as the TAMS, and thus the work of \citet{Georgy2013grid}
does not rule out other factors influencing the star's evolution towards
becoming a red supergiant. The variation in the
evolutionary tracks of the models in Fig.~\ref{teff_he4_plots} is
also due to the differences in the duration of the convective H-shell
from model to model.  The models
spending the largest fraction of their helium-burning lifetimes in the
bluer, hotter side of the HRD are those with the longest convective
H-shell duration.  The H-shell efficiency is for the most part set by
the total stellar luminosity.  Thus, the lowest luminosity models will
be able to sustain a convective H-shell for longer.  The total stellar
luminosity at this time is a product of the core mass, which is in
turn a product of the extent of convective H-burning core during the
main sequence.  The KEPLER models have the smallest convective
hydrogen-burning cores (Fig.~\ref{hcore_t_plots} and
Table~\ref{cores_table}) and thus spend the largest fraction of their
helium-burning lifetimes in the bluer, hotter side of the HRD.

\subsubsection{Mass loss}
\begin{figure}
  \centering
  \includegraphics[width=1.\linewidth,clip=True,trim=0mm
    9mm 0mm 20mm]{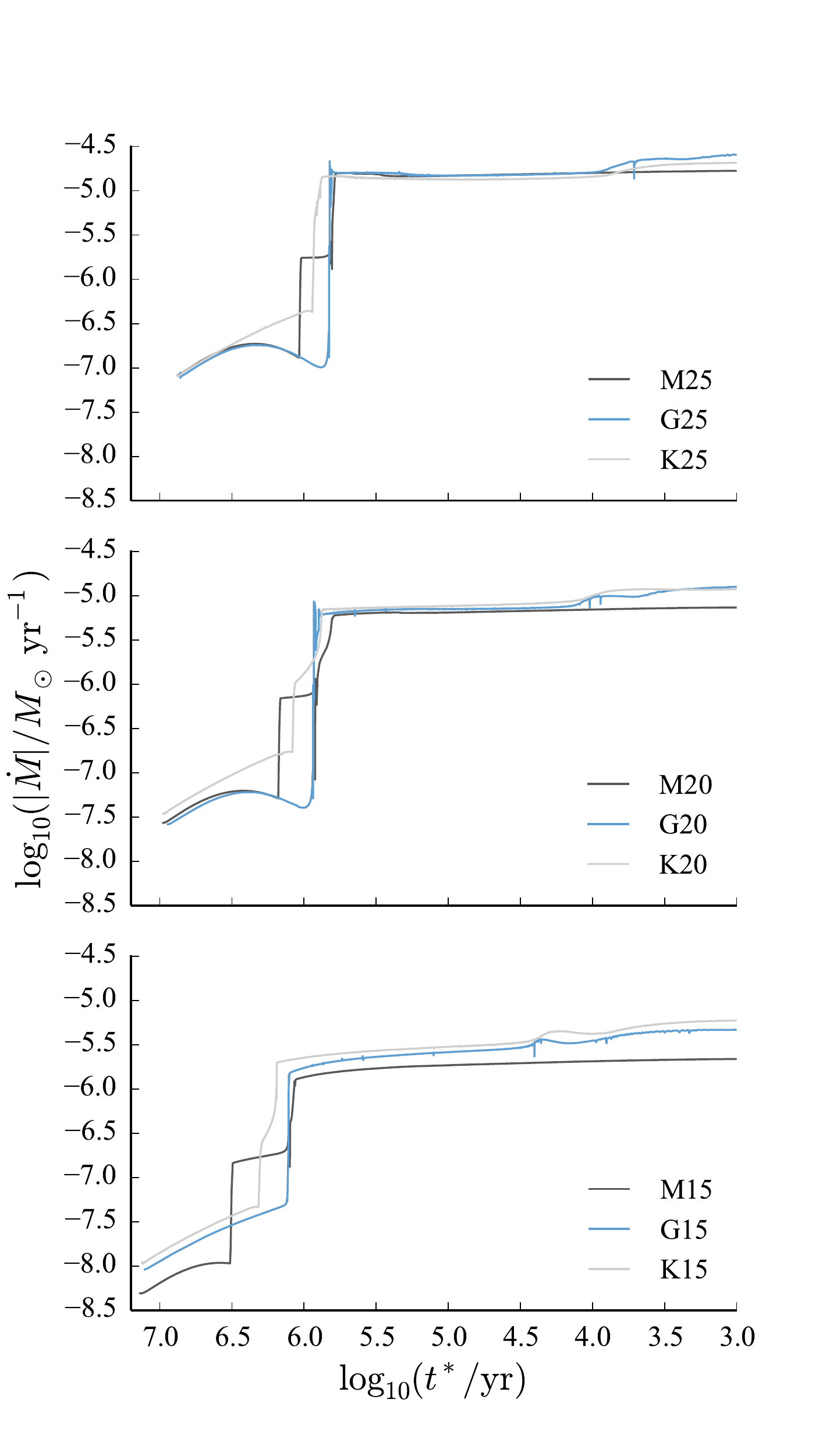}
  \caption{Mass loss rates (in $M_\odot\,\mathrm{yr}^{-1}$)
    as a function of
    $\mathrm{log}_{10}(t^*/\mathrm{yr})$, where $t^*$ is the time left
    until core collapse.}
  \label{mdot_t_plots}
\end{figure}
The mass-loss rates of our models are shown as a function of time
until core collapse in Fig.~\ref{mdot_t_plots}.  As we have discussed
in Section~\ref{mass_loss_section}, the rate of mass loss from the
stellar surface depends strongly on the position of the star in the
Hertzsprung-Russell diagram (i.e., its effective temperature and
luminosity). Mass loss is generally stronger for higher luminosity and
lower effective temperature, meaning that $15$--$25\msun$ stars
experience modest mass loss during the main sequence, strong mass loss
during the red supergiant (RSG) phase, and an intermediate rate of
mass loss on the blue supergiant (BSG) phase.

Despite mass loss being modest during the main sequence, we still find
significant discrepancies between the codes. The general shapes of the
mass loss curves as a function of time (shown in
Fig. \ref{mdot_t_plots}) during the main sequence fall into two
categories. Those models assuming the mass loss rates of
\citet{Vink2001}---the $20\msun$ and $25\msun$ GENEC models and all of
the MESA models---experience a diminishing mass loss rate before the
end of the main sequence.  This is due to the second order polynomial
dependence on the effective temperature of the \citet{Vink2001} mass
loss prescription.  The KEPLER models (\citealt{Nieuwenhuijzen1990}
mass loss rate) and the $15\msun$ GENEC model (\citealt{deJager1988}
mass loss rate), do not experience a reduction in mass loss towards
the end of the main sequence.  The sharp increase in the mass loss
rate of the MESA models (by about a factor of $10$) during the main
sequence is caused by the transition from the hot side to the cool
side of the bi-stability limit \citep[][and references
  therein]{Vink2001}.  The transition across the bi-stability limit is
not seen in the (20\msun\ and 25\msun) GENEC models that use the same
mass loss rates (Fig.~\ref{mdot_t_plots}), even though their evolution
in the HRD (in particular, their effective temperatures) is very
similar to the MESA models (Fig.~\ref{HRDs}). This is a result of the
way in which the bi-stability limit is used in the two codes. In
GENEC, the bi-stability temperature limit $T_{\rm eff}^{\rm jump}$ is
calculated using equations 14 and 15 of \citet{Vink2001}, and the mass
loss rate is calculated based on whether the effective temperature is
hotter or cooler than $T_{\rm eff}^{\rm jump}$.  In MESA, the
transition in mass loss rate from the hot side to the cool side of the
bi-stability limit is smoothed for effective temperatures in the range
$-100\,{\rm K} < T_{\rm eff}-T_{\rm eff}^{\rm jump} < 100\,{\rm
  K}$. The resulting mass loss rate for a star with effective
temperature in this range is a weighted average of the hot-side and
cool-side rates.

The total stellar mass is relatively fixed following central helium
depletion because of the rapidity of the remaining evolutionary phases
compared with the mass loss rate.  Two striking features are that
firstly, the total mass of K15 is significantly lower than in G15 or
M15, which agree very well (Table~\ref{cores_table}).  This is due to
higher mass loss rates during the RSG phase as a result of the
significantly lower effective temperature.  The lower effective
temperature in the KEPLER model is explained by the different
integration method for the envelope, which yields effective
temperatures that are too low.  This also affects the effective
temperature of the 20\msun\ model to a smaller extent.  Secondly, the
amount of mass retained in the envelope in M20 is significantly higher
than in G20 or K20.  This is due to the combined effect of two
influencing factors.  Firstly, the model spends some of its
helium-burning lifetime as a blue supergiant with higher effective
temperature and lower luminosity.  Secondly, the total helium-burning
lifetime of M20 is shorter than in G20 or K20 and thus less mass is
lost.  Lower $T_{\rm eff}$ and higher $L$ can be seen to result in
stronger mass loss (the increase in mass loss during the RSG stage)
for example in the $15\msun$ models at $\log_{10}(t^*/{\rm yr})\approx
6$ in the lower panel of Fig.~\ref{mdot_t_plots}).

\begin{table}
\begin{center}
\begin{tabular}{r c c c}
\hline \hline
Code & $15\msun$ & $20\msun$ & $25\msun$ \\
\hline \hline
\multicolumn{4}{c}{stellar structure networks}\\
\hline
GENEC & 0.499 & 0.424 & 0.372 \\
KEPLER & 0.288 & 0.261  & 0.235 \\
MESA & 0.433 & 0.391 & 0.360 \\
\hline
\multicolumn{4}{c}{post-processing network}\\
\hline
GENEC & 0.509 & 0.436 & 0.384 \\
KEPLER & 0.347 & 0.321  & 0.293 \\
MESA & 0.410 & 0.408 & 0.318 \\
\hline \hline
\end{tabular}
\caption{C/O ratios at the end of core helium burning for all the
  stellar models, both from the networks that were used to compute
  stellar structure (upper half) and from MPPNP (lower half).}
\label{CO_table}
\end{center}
\end{table}

% !TEX root = ./ms.tex

\section{Nucleosynthesis up to end of core helium burning}
\label{nucleo}

After discussing the differences in the structure of the models
calculated with the three stellar evolution codes, in this section, we
discuss the nucleosynthesis results obtained until the end of core
helium burning.  The nucleosynthesis was calculated using the same
post-processing code for all models, adopting the same nuclear
reaction network and rates (see Section~\ref{sec:MPPNP}).  In this
study, we focus on the evolution of the surface composition
(\ref{surface_composition_sect}) and the weak \textsl{s}-process
production during core helium burning
(\ref{core_he_burning_nuc_sect}).

\begin{figure}
\centering \includegraphics[width=1.\linewidth,clip=True,trim=0mm 9mm
  0mm 20mm]{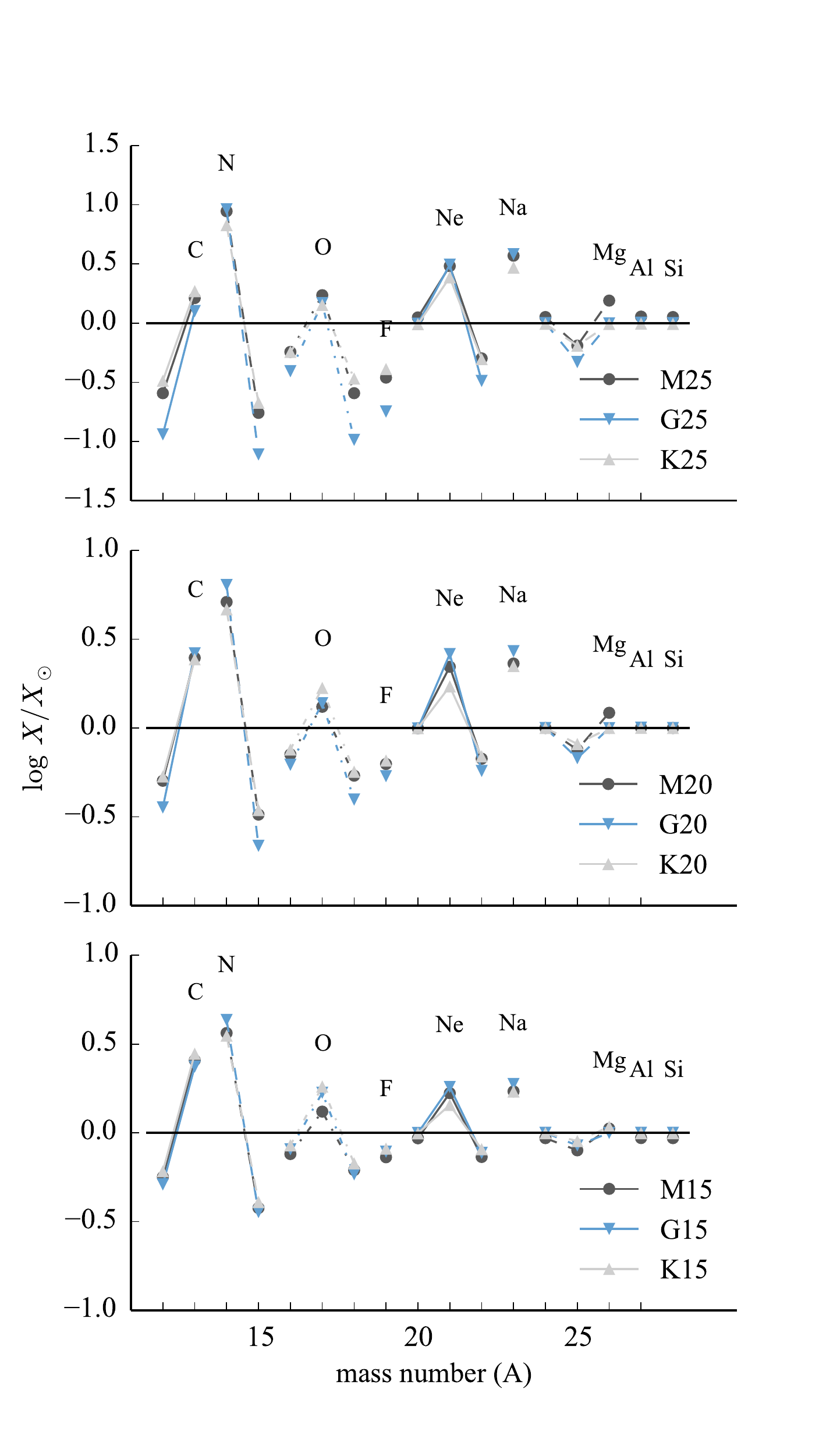}
\caption{Surface overabundances ($X_i/X_\odot$) at the end of
  the core helium-burning phase.}
\label{fig:final_surf_SJ}
\end{figure}

\subsection{Envelope and surface composition}
\label{surface_composition_sect}
Following the main sequence, the envelopes of massive stars become
convective, reaching down and dredging up material that has been
processed by the hydrogen-burning core and shell.  This dredge-up
results in a CNO processing signature observed in the surface
composition that persists until the pre-supernova stage.  This
signature includes a higher concentration of, e.g., \iso{13}C,
\iso{14}N and \iso{23}{Na}, and a lower abundance of, e.g., \iso{12}C,
\iso{15}N and \iso{16}O (Fig.~\ref{fig:final_surf_SJ}).  This
signature is stronger in more massive stars due to the more extended
(in time) convective H-burning shell.
Since it is mainly determined by nuclear reactions, this CNO
processing signature is qualitatively extremely similar between the
three codes, showing that the use of a different code has little
impact in this context.  In general, the H-burning signature on the
surface is increasing with the initial mass of the star across the
three codes.  The small quantitative differences are the following.
GENEC shows the strongest enrichment signature with the highest
enrichment for \iso{14}N and \iso{23}{Na}.  This is accompanied with a
weaker enrichment in \iso{13}C---which is a signature of incomplete
CNO processing---and the strongest depletion in \iso{12}C, \iso{15}N
and \iso{16}O, whereas KEPLER (Ledoux with semiconvection) shows the
smallest enrichment.

\subsection{Nucleosynthesis up to the end of core He-burning}
\label{core_he_burning_nuc_sect}
During the convective core helium-burning, the important reactions
competing for the economy of the $\alpha$ particles are the
$3\alpha\rightarrow\,\miso{12}{C}$,
\iso{12}{C}($\alpha,\gamma$)\iso{16}{O} and the \iso{22}{Ne}+$\alpha$
reactions, although the latter two from this list are only activated at the end of
core helium burning \citep[e.g.,][]{raiteri:91a}.  The growth of the
convective helium core and the treatment of mixing across its
convective boundary are critical factors in determining how much fresh
helium is introduced into the core and at what rate.  Introducing
fresh helium into the convective core towards the end of the
helium-burning phase when the helium abundance is becoming low is of
particular importance.  In this condition, the triple-$\alpha$
reaction is less of a competitor for the other two reactions in
consuming the freshly introduced helium and two things happen: the C/O
ratio decreases (see Section~\ref{early_cores} and
Table\,\ref{CO_table}) and the neutron exposure (or the total amount
of neutrons made by the $\miso{22}{Ne}(\alpha,n)\miso{25}{Mg}$
reaction) increases.

As described in Section~\ref{early_cores}, the convective core grows
in mass over the duration of the helium-burning phase because of the
growth of the helium core due to hydrogen shell burning and the
increased opacity due to the conversion of helium into carbon and
oxygen.  The penetrative overshooting in GENEC introduces new
helium-rich composition into the convective helium-burning core when
it first develops, after which the composition is chemically
homogenised by convection and uniformly depleted.  The introduction of
fresh, helium-rich material into the convective core can thus be
considered only a result of the growth of the extent of the convective
core and not the overshooting.  In MESA on the other hand, fresh alpha
particles are continuously mixed into the convective core from the
radiative layers above.  This is because of the diffusive treatment of
convective boundary mixing in MESA and occurs even when the abundance
of helium in the core is low, i.e., towards the end of the core
helium-burning phase.  Something similar occurs in the KEPLER models
because semiconvection above the convective core is treated as a
diffusive process.  Also affecting the size of the convective core
towards the end of core helium burning is the increase of the opacity
due to the rising amounts of C and O.  The radiative gradient (in
particular, at the edge of the convective core) will increase as a
result of this opacity increase (see Section~\ref{early_cores} and the
discussion of core breathing).  In GENEC and MESA, which consider only
the Schwarzschild criterion for convection, the convective core will
engulf material from the overlying layers when this situation
rises.  In KEPLER on the other hand, which considers the Ledoux
criterion and semiconvection, the stabilising effect of the mean
molecular weight gradient across the interface of the convective core
is considered.  The mixing is thus semiconvective rather than
convective, meaning that it would operate on the longer, secular
timescale instead of the convective timescale.

\begin{figure}
\centering \includegraphics[width=1.\linewidth,clip=True,trim=0mm 9mm
  0mm 20mm]{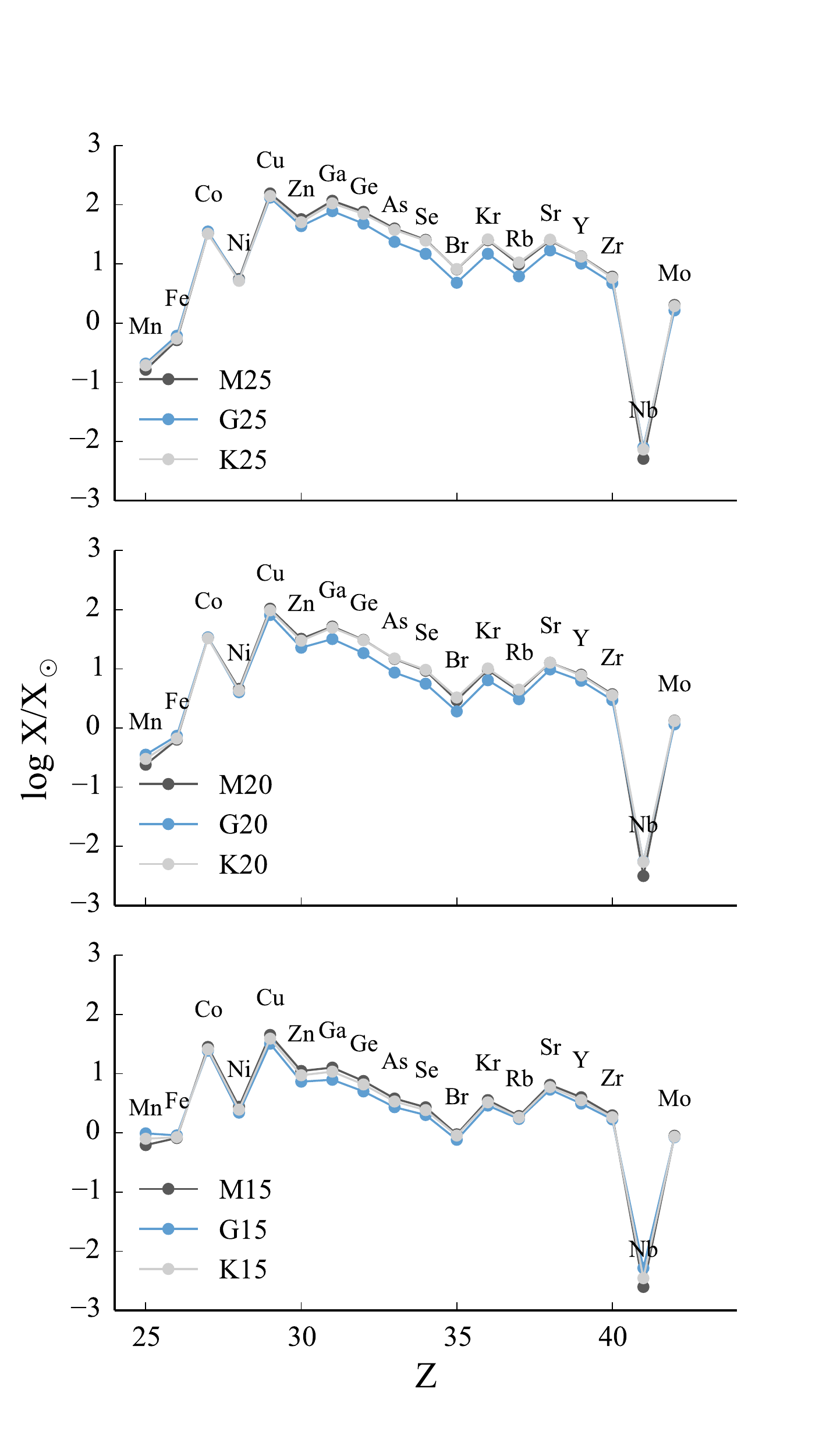}
\caption{Overabundances ($X/X_{\rm ini}$, where $X_{\rm ini}=X_\odot$)
  in the inner 1\msun\
  at the end of the core helium-burning phase.}
\label{fig:hecore_15}
\end{figure}

\begin{figure}
\centering
\includegraphics[width=0.5\textwidth]{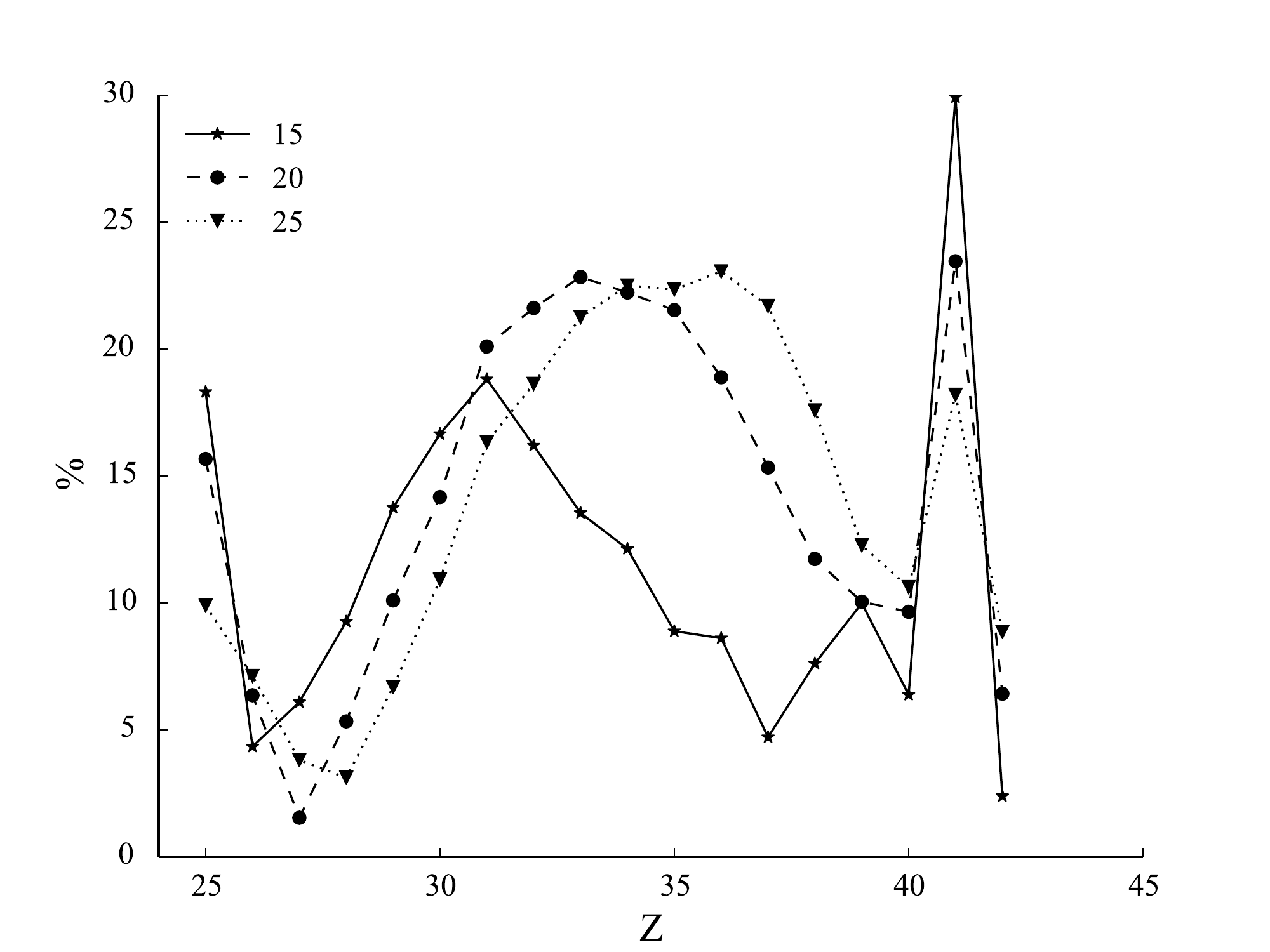}
\caption{Standard deviations (as percentages) between the three
  stellar evolution codes (MESA, KEPLER and GENEC) for the elemental
  overproduction factors ($X/X_{\rm ini}$, where $X_{\rm
    ini}=X_\odot$) in the helium core.}
\label{percent_weaks}
\end{figure}

In Fig.~\ref{fig:hecore_15} the \textsl{s}-process distributions in
the helium-depleted core from all of the models are shown at the end
of the core helium-burning phase.  The models agree well with the
GENEC models always slightly underproducing compared with the KEPLER
and MESA models.  All three codes however produce results in the weak
\textsl{s}-process region within $25\,\%$ (Fig.~\ref{percent_weaks}).
The differences are mildest for the $15\msun$ models, with the three
codes providing overproduction factors in the weak $s$-process region
that are consistent to within $20\,\%$.
Note that for all three initial masses, our results from different
stellar evolution codes show variations well within the impact due to
nuclear reaction rate uncertainties \citep{Pignatari2010,West2013}.

A comparison of the weak \textsl{s}-process distribution at the end of
the He core was already provided by \cite{kaeppeler:94}, where the
results from the codes FRANEC and G{\"o}ttingen were discussed for
massive stars with a range of initial masses.  In that work the
differences were much larger than what we obtained here.  The
production factors of \iso{80}{Kr} in the work by \cite{kaeppeler:94}
show a spread between factors of $3$ (for the 30\msun\ models) and
$25$ (for the 15\msun\ models).  In our models, the \iso{80}{Kr}
production factors agree within a factor of $2$ for a given initial
stellar mass (Table~\ref{Kr_table}).  The standard deviations between
the three codes for each initial mass and element is shown in
Fig.~\ref{percent_weaks} for the elemental production factors in the
helium core.  Part of the reason for the spread found by
\citet{kaeppeler:94} is that while the codes were using the same
$\miso{22}{Ne}+\alpha$ rates, different nucleosynthesis networks were
used for the simulations.  Here we obtain a better consistency (within
$25\,\%$ for the \textsl{s}-process region) for models with a similar
range of stellar masses, giving a brighter view from the comparison.

\begin{table}
\begin{center}
\begin{tabular}{r c c c}
\hline \hline
Code & $15\msun$ & $20\msun$ & $25\msun$ \\
\hline \hline
GENEC & 11.2 & 29.9 & 76.3 \\
KEPLER & 13.7 & 52.1  & 137.5 \\
MESA & 15.3 & 49.3 & 136.6 \\
max/min & 1.37 & 1.74 & 1.80 \\
average & $13.4\pm1.7$ & $43.8\pm9.9$ & $116.8\pm28.6$ \\
& ($13\,\%$) & ($23\,\%$) & ($24\,\%$) \\
\hline \hline
\end{tabular}
\caption{\iso{80}{Kr} overproduction factors at the end of core helium burning.}
\label{Kr_table}
\end{center}
\end{table}

In general, for different initial masses and codes Cu has the highest
production efficiency by the weak \textsl{s}-process at the end of the
He core \citep[e.g.,][]{Pignatari2010}.  On the other hand, the
production of heavier \textsl{s}-process elements like Ga and Ge
depends on the initial mass of the star and on the code, i.e., on the
amount of neutrons made by the \iso{22}{Ne}($\alpha$,n)\iso{25}{Mg}.

\begin{figure}
\centering \includegraphics[width=1.\linewidth,clip=True,trim=0mm 9mm
  0mm 20mm]{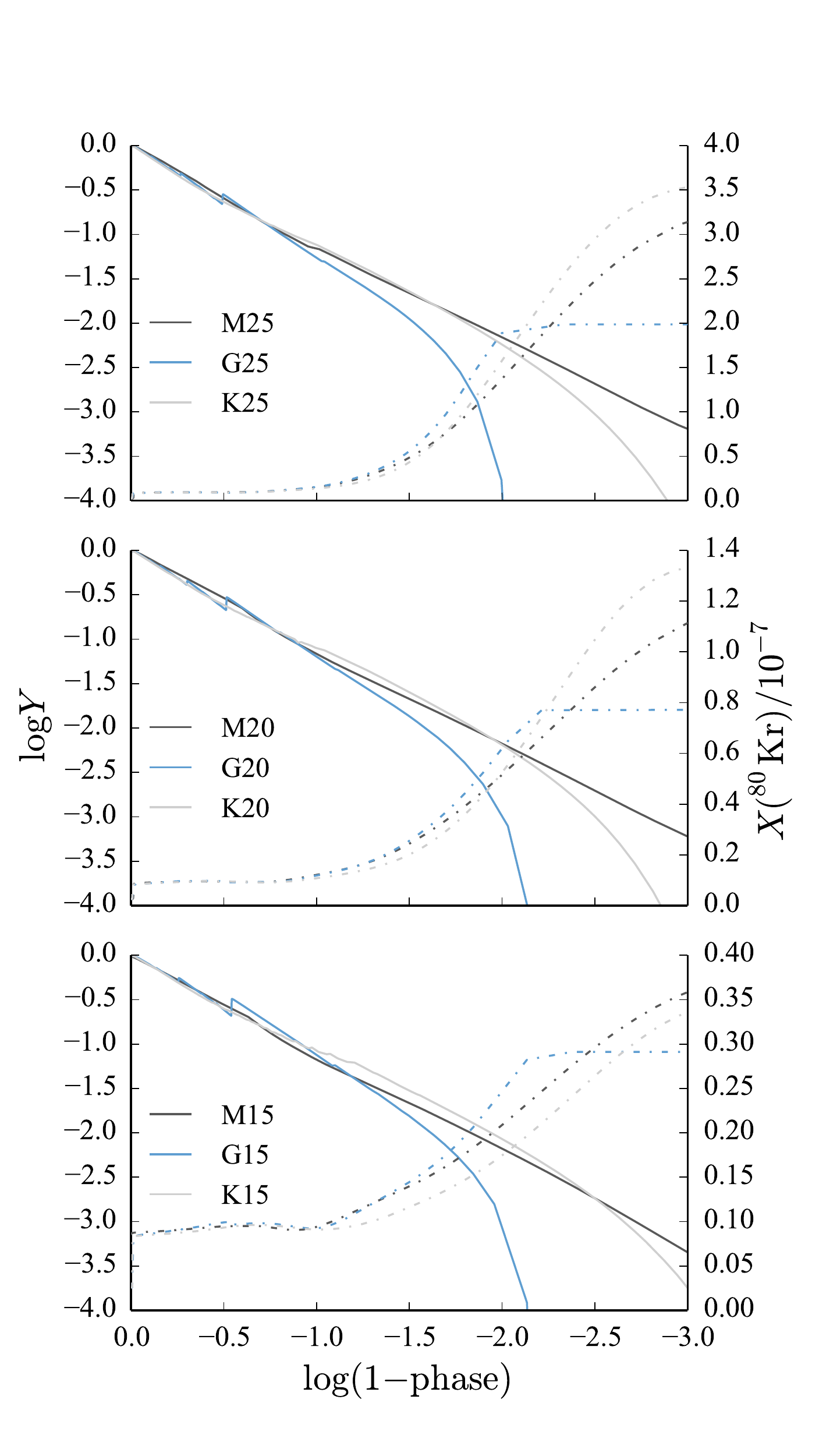}
\caption{Central \iso{4}{He} (solid lines) and \iso{80}{Kr} ($s$-only;
  dot-dashed lines) abundances during the core helium-burning phase.
  The $x$-axis is $\log(1-{\rm phase})$, where ${\rm phase}=0$ at
  helium ignition (left side of the plot) and $1$ at helium depletion
  (right side of the plot).}
\label{fig:He_time}
\end{figure}

\begin{table}
\begin{center}
\begin{tabular}{r c c c}
\hline \hline
Model & $X_{\rm C}(\miso{22}{Ne})~/~10^{-2}$ & $X_{\rm C}(\miso{56}{Fe})~/~10^{-3}$ & $T_{\rm C}~/~10^8$ K \\
\hline \hline
G15 & 1.506 & 0.704 & 3.23 \\
K15 & 1.384 & 0.637 & 3.11 \\
M15 & 1.329 & 0.588 & 3.11 \\
\hline
G20 & 1.100 & 0.457 & 3.40 \\
K20 & 0.940 & 0.409 & 3.28 \\
M20 & 0.900 & 0.377 & 3.29 \\
\hline
G25 & 0.768 & 0.331 & 3.55 \\
K25 & 0.615 & 0.306 & 3.39 \\
M25 & 0.544 & 0.272 & 3.36 \\
\hline \hline
\end{tabular}
\caption{Central abundance (mass fraction) of \iso{22}{Ne} and central
  temperature at the end of core helium burning.}
\label{Ne22_Tc_table}
\end{center}
\end{table}

\begin{figure}
\centering \includegraphics[width=1.\linewidth,clip=True,trim=0mm 7mm
  0mm 20mm]{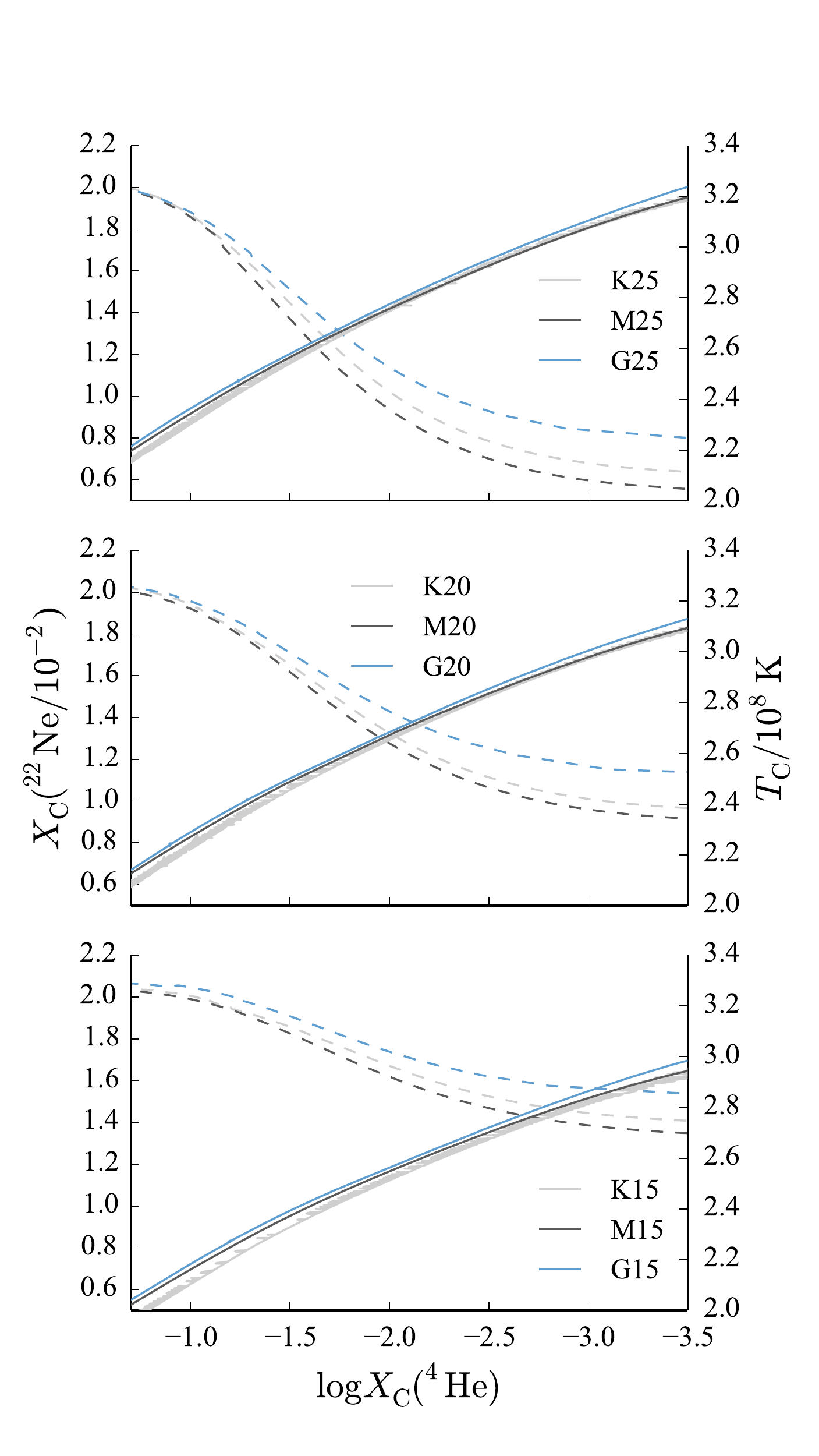}
\caption{Central \iso{22}{Ne} abundances (dashed lines) and central
  temperatures (solid lines) during the core helium-burning phase.}
\label{fig:Ne_time}
\end{figure}

In Fig.~\ref{fig:He_time} the evolution of the \iso{4}{He} and
\iso{80}{Kr} abundances in the convective He-burning core are shown
with respect to phase (where ${\rm phase}=0$ at helium ignition and 1
at helium depletion).  \iso{80}{Kr} is defined as an $s$-only
isotope, but it also receives a relevant explosive contribution, from
the classical \textsl{p}-process \citep[e.g.,][]{arnould:03}, from
neutrino-driven wind nucleosynthesis components
\citep[e.g.,][]{froehlich:06,farouqi:10,arcones:11} and eventually
from $\alpha$-rich freezout ejecta \citep{woosley:92}.  Its production
in \textsl{s}-process conditions strongly depends on the branching in
the neutron-capture path at \iso{79}{Se} \citep{klay:88}.  We focus in
particular on \iso{80}{Kr} in Fig.~\ref{fig:He_time} because there are
detailed results for this nucleus reported in the study by
\citet{kaeppeler:94} to which we will have compared the results of our
calculations.  The increase of \iso{80}{Kr} starts during the last
10\% of the He-burning lifetime ($\log(1-{\rm phase})=-1$ in
Fig.~\ref{fig:He_time}).  Since \iso{22}{Ne} is not fully consumed
during the He core, \iso{80}{Kr} increases until the reservoir of
$\alpha$-particles is exhausted.  Furthermore, the amount of
\iso{80}{Kr} made increases with the initial mass of the star (note
the different $y$-axis scales on the right hand side of
Fig.~\ref{fig:He_time}).  This means that the \textsl{s}-process
efficiency increases with the initial mass of the star because of the
higher central temperatures and the more efficient
\iso{22}{Ne}($\alpha$,n)\iso{25}{Mg} activation, as already discussed
by \cite{prantzos:90}.  This trend of increasing \textsl{s}-process
efficiency with increasing initial stellar mass is also shown in
Fig.~\ref{fig:hecore_15}. The evolution of the central temperature and
\iso{22}{Ne} abundance during the core helium burning phase is shown
in Fig.~\ref{fig:Ne_time}. The values of these two quantities at the end
of the core helium burning phase is given together with the abundance
of \iso{56}{Fe} in Table~\ref{Ne22_Tc_table}.

The three codes that were used to produce these stellar models assumed
different formulations for the treatment of convective boundary mixing
(CBM) above the convective He core.  These are: penetrative
overshooting (GENEC), fast semi-convection (KEPLER) and exponentially
decaying diffusive overshooting (MESA).  These formulations are
described in more detail in Section~\ref{conv_ovr_section}.  The
impact of overshooting on the weak \textsl{s}-process was considered
by, e.g, \citet{Langer1989} and \citet{Pumo2010}, showing that in
general the \textsl{s}-process production increase with the
overshooting efficiency, due to the larger reservoir of \iso{4}{He}.
Fig.~\ref{fig:He_time} shows that the \textsl{s}-process production
(represented by \iso{80}{Kr}) is drawn-out by continuous replenishment
of $\alpha$-particles in the KEPLER and MESA models due to their
time-dependent, diffusive CBM treatments.  Nevertheless, we show
something new with these results compared to \citet{Langer1989}: three
alternative CBM formulations each with their own choice of parameters
are providing consistent (overproduction of the weak
\textsl{s}-process elements agreeing to within $25\,\%$)
\textsl{s}-process results at the end of the He core.

% !TEX root = ./ms.tex

\section{Discussion and concluding remarks}
\label{discussion}

We have compared the structural evolution of stellar models from the
GENEC, KEPLER and MESA stellar evolution codes. We chose models with
initial masses $15\msun$, $20\msun$, and $25\msun$ at $Z=0.02$, which
is arguably the most studied massive star mass range.  The models were
analysed from the zero-age main sequence to the end of the core
helium-burning stage.  We have computed the full ($1088$ species)
nucleosynthesis of all nine models in a post-processing mode using the
NuGrid code MPPNP, in which a mixing (diffusion) step is performed
after each network time step (i.e., the mix and burn operators are
decoupled).

During the hydrogen and helium-burning stages, the main differences in
the structure of the models can be traced back to the different
choices for the input physics. The key differences are the criterion
for convective stability: Schwarzschild (GENEC and MESA) compared to
Ledoux (KEPLER), the treatment of convective boundary mixing:
penetrative overshooting (GENEC), exponentially decaying diffusion
scheme (MESA) or semi-convection (KEPLER) as well as the mass loss
rate prescriptions and their implementation.  The size of the
convective core during the main sequence and the main sequence
lifetime is mostly influenced by the treatment of convective boundary
mixing or overshooting.  The chemical composition of the core at
helium depletion is very sensitive to the structural evolution during
core helium burning.  Particularly sensitive are the ratio of carbon
to oxygen and the weak \textsl{s}-process component.  The three codes
show quite a large spread in the C/O ratio in the helium-depleted core
owing to their different assumptions on convective boundary mixing and
their different helium-burning lifetimes.  Another factor is of course
the rates used for the key nuclear reactions.  Whereas the rates for
the $3\alpha$ and $\miso{12}{C}(\alpha,\gamma)\miso{16}{O}$ reactions
are clearly important, the C/O ratio is also mildly sensitive to the
choice of $\miso{22}{Ne}(\alpha,n)\miso{25}{Mg}$ and
$\miso{22}{Ne}(\alpha,\gamma)\miso{26}{Mg}$ rates. Of course, the C/O
ratio shows a more marked change when the $\alpha$-capture channels of
\iso{22}{Ne} are omitted completely from the reaction network.  This
strongly suggests that \iso{22}{Ne} should be included in even small
networks whose aims are solely to accurately compute the energy
generation to supplement the structural evolution of the stellar
model.

The treatment of convective boundary mixing during the main sequence
affects the hydrogen and helium profiles outside of the helium core.
The location of the star in the Hertzsprung-Russell diagram as the
star is becoming a supergiant following the main sequence is very
sensitive to these profiles \citep[see][]{Georgy2013grid}.
The KEPLER models spend less time as red supergiants
than the MESA and GENEC models.
This is because the duration of the convective hydrogen-burning shell
is longest in the KEPLER models---a result of their smaller
hydrogen-burning convective cores and hence lower luminosities.  The
main sequence lifetimes of the models from the three codes agree
within $5\,\%$ or better for a given initial mass, while the
helium-burning lifetimes show a much larger spread (up to $23\,\%$ in
the 15\msun\ models).  The core masses and total masses of the models
at the end of core helium burning show spreads with standard
deviations in the range $2\,\%$--$9\,\%$.  Particularly for the helium
and CO core masses, the standard deviation is dominated by the smaller
KEPLER cores (because of the choice of Ledoux criterion and
semiconvection compared to Schwarzschild criterion and overshooting in
the other two codes).  That being said, the CO cores show less of a
spread than the helium cores due to the longer helium-burning
lifetimes in the KEPLER models.  The spread is a little more even in
the total stellar masses for the 20\msun\ and 25\msun\ models.

The main results concerning the differing nucleosynthesis in the
models can be summarised as follows: The surface abundances show the
characteristic signature of CNO enrichment, which is stronger for
larger initial masses.  This enrichment is linked to the physical
extent and duration of the convective hydrogen-burning episode
preceding the dredge-up and always shows the strongest signature in
the GENEC models and the weakest in those of KEPLER.  We compared the
\textsl{s}-process results obtained at the end of the He core, with an
approach similar to \cite{kaeppeler:94} but using the same
post-processing nuclear network for the three sets of models.
Compared to these earlier comparison, we obtained much smaller
departures between the results from different codes.  Concerning the
\textsl{s}-process elements between Fe and Sr, the largest differences
are obtained for the overproduction factors of Ga (in the
15\msun\ models), As (in the 20\msun\ models) and Kr (in the
25\msun\ models), with standard deviations of $17\,\%$, $23\,\%$ and
$23\,\%$, respectively.  The standard deviation of the elemental
production factor of any element between Fe and Mo in the helium core
is less than $30\,\%$.  These differences in the elemental
overproduction factors are much smaller than the impact of the present
nuclear physics uncertainties \citep[e.g.,][]{Pignatari2010}.
Therefore, the three codes can be considered to yield consistent
results.  The discussion may be more complex for the single isotopes,
where larger differences can be obtained, but overall we may conclude
that the results are consistent.  We anticipate that while the
overproduction factors look consistent at the end of the helium core,
the final production factors might be quite different due to, e.g.,
the impact of different He core sizes, the \textsl{s}-process
activation during the following convective carbon shells burning and
the final core-collapse supernova modification of the ejected \textsl{s}-process
abundances.  The amount of weak \textsl{s}-process material for a
given progenitor mass, however, may have implications for
galactic chemical evolution studies \citep[e.g.,][]{Brown2013}.

The models presented in this paper show that the three stellar
evolution codes yield consistent results, which is reassuring for the
field.  Differences in key properties of the models, like helium and
CO core masses, are traced back to the treatment of convection
\citep[see also][]{Sukhbold2013}.  The behaviour of the mixing
processes in stars remains the uncertainty of primary concern in
stellar modelling.  Better constrained prescriptions are thus
necessary to improve the predictive power of stellar evolution models.
Multi-dimensional hydrodynamic simulations and asteroseismological
observations will hopefully provide the necessary constraints to
reduce the current uncertainties in the coming decade.

\section*{Acknowledgments}

NuGrid acknowledges significant support from NSF grants PHY 02-16783
and PHY 09-22648 (Joint Institute for Nuclear Astrophysics, JINA).
R.H. and C.G. acknowledge the support from Eurocore project
Eurogenesis and ERC Starting Grant No.~306901.  R.H. acknowledges
support from the World Premier International Research Center
Initiative (WPI Initiative), MEXT, Japan.  M.P. acknowledges the
support from the Ambizione grant of the SNSF and the SNF grant
(Switzerland).  A.H. acknowledges the support by NSF through grant
AST-1109394, by US DOE grants FC02-09ER41618 (SciDAC),
DE-FG02-87ER40328, and by an ARC Future Fellowship (FT120100363).

\bibliographystyle{mn2e_fixed}

\bibliography{references}

\label{lastpage}

\end{document}